\pdfoutput=1
\RequirePackage{rotfloat}
\documentclass[fleqn,usenatbib]{mnras}

\usepackage[T1]{fontenc}
\usepackage{ae,aecompl}

\usepackage{graphicx}	
\usepackage{amsmath}	
\usepackage{amssymb}	
\hypersetup{draft}
\usepackage{threeparttable}
\usepackage{rotfloat}
\usepackage{ulem}
\usepackage{enumerate}
\usepackage{pdflscape}

\title[Cluster Galaxy Rotational Profiles]{The Rotational Profiles of Cluster Galaxies}

\author[L. E. Bilton et al.]{Lawrence E. Bilton,$^{1}$\thanks{E-mail: \href{mailto:l.bilton@2016.hull.ac.uk}{l.bilton@2016.hull.ac.uk}}
Matthew Hunt,$^{1}$
Kevin A. Pimbblet$^{1}$
\newauthor
and Elke Roediger$^{1}$
\\
$^{1}$E.A. Milne Centre for Astrophysics, The University of Hull, Cottingham Road, Kingston upon Hull, HU6 7RX, UK\\
}

\date{Accepted XXX. Received YYY; in original form ZZZ}

\pubyear{2019}

\begin{document}
\label{firstpage}
\pagerange{\pageref{firstpage}--\pageref{lastpage}}
\maketitle

\begin{abstract}
We compile two samples of cluster galaxies with complimentary hydrodynamic and N-body analysis using FLASH code to ascertain how their differing populations drive their rotational profiles and to better understand their dynamical histories.
We select our main cluster sample from the X-ray Galaxy Clusters Database (BAX), which are populated with Sloan Digital Sky Survey (SDSS) galaxies.
The BAX clusters are tested for the presence of sub-structures, acting as proxies for core mergers, culminating in sub-samples of 8 merging and 25 non-merging galaxy clusters.
An additional sample of 12 galaxy clusters with known dumbbell components is procured using galaxy data from the NASA/IPAC Extragalactic Database (NED) to compare against more extreme environments.
BAX clusters of each sample are stacked onto a common RA-DEC space to produce rotational profiles within the range of $0.0 - 2.5$ $r_{200}$.
Merging stacks possess stronger core rotation at $\lesssim 0.5 r_{200}$ primarily contributed by a red galaxy sub-population from relaxing core mergers, this is alongside high rotational velocities from blue galaxy sub-populations, until, they mix and homogenise with the red sub-populations at $\sim r_{200}$, indicative of an infalling blue galaxy sub-population with interactive mixing between both sub-populations at $\gtrsim r_{200}$.
FLASH code is utilised to simulate the merger phase between two originally independent clusters and test the evolution of their rotational profiles.
Comparisons with the dumbbell clusters leads to the inference that the peculiar core rotations of some dumbbell clusters are the result of the linear motions of core galaxies relaxing onto the potential during post second infall.
\end{abstract}




\section{Introduction}
\label{sec:intro}

Galaxy clusters are large and dense realms in space which anistropically coalesce along the convergence of independent filaments through hierarchical merger events, resulting in the induction of random motions in their member galaxies (e.g. \citealt{Bond1996,Springel2005,Kravtsov2012}). 
These large collections of matter are home to strong gravitational potentials that cause the further perturbation of galaxies from the Hubble flow \citep{Regos1989}.
As a result, galaxy clusters seemingly play host to environmental effects that are pivotal in the understanding of the evolution of galaxies through an assumption of fixed stellar mass: the transition from late-type to early-type galaxies towards the cluster's centre with the morphology-density relation (e.g. \citealt{Oemler1974,Dressler1980,Postman1984}); the observed bimodality of the colour-density relation \citep{Hogg2003,Hogg2004}; the consistent decrease in the fraction of star forming galaxies in cluster cores (e.g. \citealt{Lewis2002,Gomez2003,Bamford2009,Linden2010}); a galaxy infalling onto a cluster potential experiencing ram-pressure stripping due to interacting with the intracluster medium (ICM) (e.g. \citealt{Gunn1972,Sheen2017,Poggianti2017}). 

The hierarchical nature of galaxy cluster formation lends itself to the existence of physical substructures \citep{Geller1982,Dressler1988}.
Therefore, the aforementioned environmental effects on galaxy evolution can be scaled down to the smaller substructure environments within a cluster.
We can use the presence and strength of this sub-structuring within the cluster to delineate differing environments (i.e. merging or non-merging).
The substructures that reside at larger radii from the cluster centre are smaller galaxy groups that cause \textquoteleft pre-processing' \citep{Berrier2009,Bahe2013}; smaller-scale premature evolution of galaxies due to localised galaxy-galaxy interactions (see \citealt{Moore1999}).
Pre-processing is considered to be a common occurrence in order to account for the swift changes in star formation and colour fractions as galaxies transition from the field (e.g. see \citealt{Haines2015,Bilton2018}).

Perhaps one of the striking features of many galaxy clusters is the presence of overtly bright giant early-type galaxies, commonly with an extended diffuse region, that lie within the dynamical centres of their host cluster \citep{Quintana1982}, otherwise known as the brightest cluster galaxy (BCG). 
The formation mechanism for BCGs has been key point of contention.
One such model is galactic cannibalism \citep{Ostriker1975}, whereby galaxies infall and accumulate at the bottom of the potential well through dynamical friction.
An alternatively favoured model is rapid hierarchical galaxy-galaxy merging into an ensemble of sub-groupings of galaxies of similar size prior to collapse onto the bottom of the potential \citep{Merritt1985}.
Testing of these models has often yielded mixed results; galactic cannibalism is deemed too slow in order to build a BCG within a reasonable timescale with the observed luminosities (e.g. \citealt{Lauer1988,Dubinski1998}); hierarchical galaxy merging events alone do not assemble enough sub-groupings with calculations to our current epoch \citep{Collins2009}.
However, despite these shortfalls, there is evidence for clusters to have had merger events over their histories with the observations of BCGs with multiple cores (e.g. \citealt{Oegerle1992,Laine2003}).
There is also convincing evidence of core-core pre-merger; on-going merger; post-merger activity between two originally independent potential wells, with high peculiar velocities of BCGs, that indicate perturbations from their original geometric and kinematic centres (e.g. \citealt{Quintana1996,Smith2005a,Pimbblet2006,Shan2010,Lakhchaura2013,Caglar2017}).
These systems with multiple-core BCGs are sometimes known as \textquoteleft Dumbbell Clusters'.

If we assume these dumbbell clusters arise from two originally independent sub-clusters interacting off-axially, then the strength and presence of their resultant momenta could leave an imprint onto their line-of-sight velocities, producing some sort of \textquoteleft global cluster rotation' \citep{Ricker1998}.
Due to the apparent random motions of cluster galaxies, the idea of galaxy clusters supported by rotational energies was excused for a pressure-based model. 
However, once thought indistinguishable from the cluster galaxy kinematics, there have since been several works that have observed global rotation (e.g. \citealt{Materne1983,Oegerle1992,Hwang2007,Tovmassian2015,Manolopoulou2017}).
One could argue that the source of cluster rotation is from the Universe possessing its own angular momentum and donating it onto celestial bodies during their formation \citep{Li1998,Godlowski2003,Godlowski2005}.
However, to account for the strong peculiar velocities from relatively recent histories, galaxy cluster rotation could be derived from the merging processes between two clusters \citep{Peebles1969,Ricker1998}; off-axis tidal interactions from two independent deep potential wells.
Observations of such events/relics would only be pragmatic by observing the ICM due to the high collisional probability of particles that produce X-rays, where the more sparse cluster galaxies are found to be collisionless on equal timescales; global angular momentum observed via the galaxies is transient \citep{White1995,Roettiger1997,Roettiger2000}.
Indeed, recent simulation studies show how the ICM could be used to determine bulk cluster rotation dynamics (see \citealt{Baldi2017,Baldi2018}).
In addition, there is the very sensible notion that the accretion of mass through the filaments during cluster assembly is the primary driver of momentum donation to these systems (see \citealt{Song2018}).
One immediate method that could be used to infer a cluster's global rotation is the use of a geometrical technique known as \textquoteleft perspective rotation'; peculiar motion measurements taken from the mean radial velocities to determine the transverse motion by artificially rotating the galaxy cluster on the plane of the sky \citep{Feast1961}.

Therefore, in this paper we aim to establish whether or not the global rotational dynamics of clusters correlate with their sub-populations and if the presence of a dumbbell BCG core imprints a cluster evolutionary mechanism onto the global rotational profile.
This will be achieved by utilising a \textquoteleft perspective rotation' technique, which infers the presence of cluster rotation through the comparative radial velocity differences between two semi-circles divided by the cluster centre (see \citealt{Manolopoulou2017}; MP17 hereafter).
We present a complementary suite of galaxy data from the Sloan Digital Sky Survey (SDSS) and NASA/IPAC Extragalactic Database (NED) to form a membership of clusters into two samples of those that do, and do not, host a dumbbell nucleus.
An elaboration on how these data were acquired can be found in section \ref{sec:data}. 
An explanation of the methods used to output our global rotation profile analysis, along with our results, are outlined in section \ref{sec:cree}.
In addition to our array of observational data, we utilise 3D hydrodynamics and N-body simulations to determine the impact of idealised binary cluster mergers on global rotational profiles.
The comparison of our observational rotational analysis with our 3D hydrodynamic and N-body simulations are elaborated in section \ref{sec:sims}.
Concluding with a discussion and summary of our findings in this work with section \ref{sec:smmy}.

Throughout this work we assume a flat $\Lambda$CDM model of cosmology with $\Omega_{M} = 0.3$, $\Omega_{\Lambda} = 0.7$, $H_{0} =$ $100h$ km s$^{-1}$ Mpc$^{-1}$, where $h = 0.7$.

\section{The Data Suite}
\label{sec:data}

We compile two samples of clusters to allow for a more comprehensive study of the affects of cluster rotation; a sample of bright X-ray selected clusters utilising the X-ray Cluster Database (BAX; \citealt{Sadat2004}), a curated repository linking X-ray data from multiple instrumental sources; a sample of dumbbell clusters catalogued by \cite{Gregorini1992,Gregorini1994}
Each X-ray selected cluster is then built from the BAX centre and defined by galaxies from SDSS Data Release 8 (DR8; \citealt{Aihara2011}) cross-matched with the MPA-JHU Value Added Catalogue \citep{Kauffmann2003,Brinchmann2004,Tremonti2004}.
The dumbbell clusters are assembled via the procurement of NED galaxies that lie within 30 arcminutes of the NED-defined centres of the 12 dumbbell clusters, similarly to the work presented by \cite{Pimbblet2008}.


We initialise our cluster sample with the BAX database by employing an X-ray luminosity range of $1<L_{X}\leq20$ $\times10^{44}$ ergs$^{-1}$ that lie within the redshift range of $0.0<z\leq0.15$.
These limits ensure we are selecting the most massive clusters; that we garner a significant number of clusters; sampling across a variety of dynamical states in $z$-space from a finite epoch range.
Once parsed through BAX, the applied limits provide an initial sample size of 481 clusters.
For each of the BAX clusters an initial radial limit of $\leq10$ Mpc $h^{-1}$ is applied to DR8 galaxies from the cluster centre, calculated with our outlined cosmology \citep{Wright2006}. 
The cluster sample is iterated through to have their global mean recession velocities ($\overline{cz}_{\text{glob}}$) and velocity dispersions ($\sigma_{\text{glob}}$) calculated for galaxies $\leq1.5$ Mpc $h^{-1}$ from the cluster centre by computing the square root of the biweight midvariance \citep{Beers1990}.
We then proceed to define the cluster boundary in velocity space as a function of projected radius $R$ with surface caustics, in concordance with the methodology of \cite{Diaferio1997,Diaferio1999}. 
Velocity limits of $\Delta V=\pm$ 1500 kms$^{-1}$  are imposed upon each cluster in the sample as a conservative threshold to increase confidence against interlopers contaminating the DR8 assembled clusters, with

\begin{equation}
    \Delta V = c \left (\frac{z_{\text{gal}}-z_{\text{clu}}}{1+z_{\text{clu}}}\right).
    \label{eq:dv}
\end{equation}

\noindent
However, it should be noted that this limit can potentially omit genuine members from those systems that are actively relaxing onto a cluster potential due to their greater dispersion of galaxies, acting as an echo from two originally independent sub-clusters coalescing onto each other.
The surface caustic profiles are then determined with the remaining galaxies for each cluster, allowing for estimations of $M_{200}$ and $r_{200}$, the cluster masses and radii for when the density is 200 times the critical density of the universe for our flat cosmology \citep{Gifford2013,Gifford2013a}.
Cluster candidates are ignored if their initial richness is $<50$ at $\leq r_{200}$, or, if they are found within the \cite{Einasto2001} supercluster catalogue to possesses overlapping structures.
The resultant cluster sample size provided by the BAX-DR8 galaxies is 33, which is found to be mass-complete at log$_{10}$($M_{*}$) $\geq10.2$.
The final compilation of BAX-defined clusters built from DR8 galaxies can be found in Table \ref{tab:bax}.

The 12 dumbbell clusters are initialised with the NED galaxies that reside within 30 arcminutes from the NED defined centres. 
The NED galaxies associated with each cluster are then run through the same process as the DR8 galaxies above.
The NED-defined clusters built from the NED galaxies, along with their calculated values, can be found in Table \ref{tab:ned}.


\begin{figure*}
    \centering
    \includegraphics[width=0.85\paperwidth]{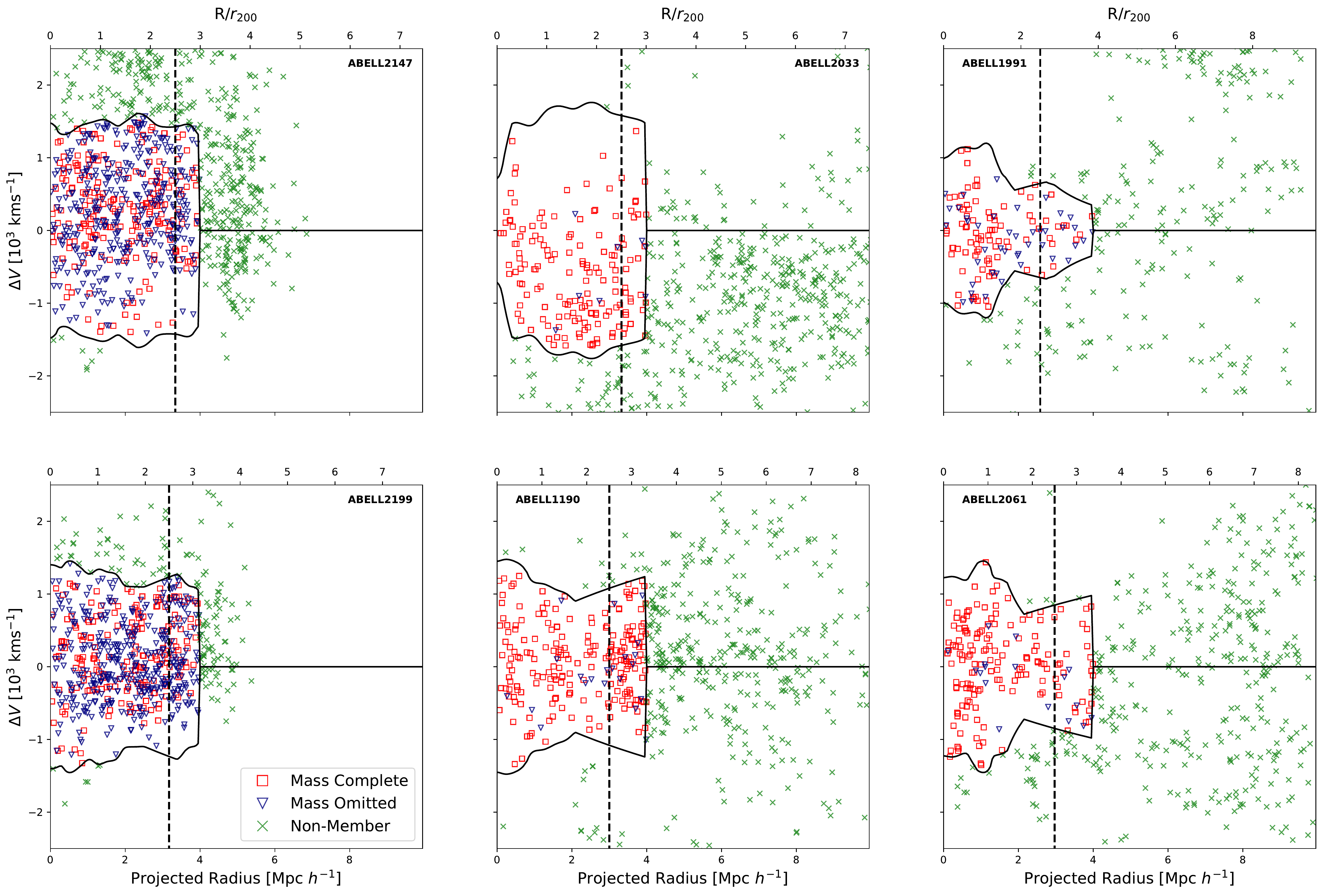}
    \caption{Example surface caustics (the black curves) from the final merging cluster sample (top row) and non-merging cluster sample (bottom row). Where the red squares represent the galaxies that make a complete sample at log$_{10}$(M$_{\ast}$) $\geq10.2$, with the blue triangles representing omitted galaxies that are at log$_{10}$(M$_{\ast}$) $<10.2$. Galaxies that lie within the surface caustics are considered to be cluster members. Here the radial velocity ($\Delta V$) with respect to the cluster's mean recession velocity is plotted against the projected radius in units of Mpc $h^{-1}$ and $R/r_{200}$. The black dashed vertical lines represent the 2.5 $R/r_{200}$ radial cut of each cluster; galaxies $\leq$2.5 $R/r_{200}$ within the caustics produce the rotational profile stacks.}
    \label{fig:caustics}
\end{figure*}



\begin{table*}
	\centering
	\caption{The mass-complete BAX cluster sample. The J2000 coordinates and X-ray luminosity values are taken from BAX. The methodology for the determination of kinematic and global rotational values can be found in sections \ref{sec:data} and \ref{ssec:MP}. $\chi^{2}_{id}/\chi^{2}_{rd}$ is the ratio of the $\chi^{2}$ statistic between ideal and random rotation curves with $P(KS)$ being the two-sample Kolmogorov-Smirnov test of significance in rotation. MP17 defines a strict criterion as $P(KS)<0.01$ and $\chi^{2}_{id}/\chi^{2}_{rd}$ $\leq0.2$, alongside a loose criterion as $P(KS)<0.01$ and $\chi^{2}_{id}/\chi^{2}_{rd}$ $\leq0.4$, for determining the presence of cluster rotation. The $P(\Delta)$ values represent the significance of sub-structuring with respect to the $\Delta$-test in equation \ref{eq:ds}. Where $P(\Delta)\ll$0.01 and $P(KS)\ll$0.01 is strongly indicative of sub-structure and rotation with values smaller three d.p.}
	\label{tab:bax}
	\begin{sideways}
	\begin{tabular}{lccccccccccr} 
		\hline
		Cluster & RA  & DEC & $L_{x}$ & $\overline{cz}_{\text{glob}}$ & $N_{r_{200}}$ & $\sigma_{r_{200}}$ & $v_{\text{glob}}$ & $\theta_{\text{glob}}$ & $\chi^{2}_{id}/\chi^{2}_{rd}$ & $P(KS)$ & $P(\Delta)$ \\
				 & (J2000) & (J2000) & ($\times 10^{44}$ erg s$^{-1}$) & (km s$^{-1}$) & & (km s$^{-1}$) & (km s$^{-1}$) & ($^{\circ}$) & & & \\
		\hline
		 & & & & & \multicolumn{1}{l}{Merging} & & & & & & \\
		\hline
		
		Abell 426 & 03 19 47.20 & +41 30 47 & 15.34$^{a}$ & 5396$\pm$62 & 106 & 831$^{+40}_{-46}$ & 271$\pm$68 & 100 & 0.26 & 0.007 & 0.010 \\
		
		Abell 1552 & 12 29 50.01 & +00 46 58 & 1.09$^{d}$ & 25782$\pm$111 & 75 & 809$^{+64}_{-84}$ & 366$\pm$129 & 260 & 0.12 & 0.083 & 0.003 \\
		
		Abell 1750 & 13 30 49.94 & -01 52 22 & 3.19$^{c}$ & 25482$\pm$95 & 70 & 726$^{+55}_{-71}$ & 512$\pm$121 & 100 & 0.08 & $\ll$0.01 & $\ll$0.01 \\
		
		Abell 1767 & 13 36 00.33 & +03 56 51 & 2.43$^{c}$ & 20985$\pm$78 & 126 & 770$^{+47}_{-58}$ & 326$\pm$90 & 320 & 0.12 & $\ll$0.01 & 0.002 \\
		
		Abell 1991 & 14 54 30.22 & +01 14 31 & 1.42$^{d}$ & 17687$\pm$61 & 57 & 535$^{+37}_{-47}$ & 287$\pm$78 & 270 & 0.07 & 0.044 & $\ll$0.01 \\
		
		Abell 2033 & 15 11 28.19 & +00 25 27 & 2.56$^{b}$ & 24582$\pm$90 & 53 & 589$^{+51}_{-69}$ & 571$\pm$100 & 300 & 0.15 & $\ll$0.01 & $\ll$0.01 \\
		
		Abell 2147 & 16 02 17.17 & +01 03 35 & 2.87$^{a}$ & 10492$\pm$48 & 95 & 688$^{+30}_{-35}$ & 226$\pm$60 & 300 & 0.15 & $\ll$0.01 & $\ll$0.01 \\
		
		Abell 2255 & 17 12 31.05 & +64 05 33 & 5.54$^{a}$ & 24283$\pm$107 & 112 & 817$^{+62}_{-80}$ & 317$\pm$92 & 60 & 0.36 & 0.011 & $\ll$0.01 \\
		
		\hline
		& & & & & \multicolumn{1}{l}{Non-Merging} & & & & & & \\
		\hline
		
		Abell 85 & 00 41 37.81 & -09 20 33 & 9.41$^{a}$ & 16488$\pm$73 & 71 & 709$^{+44}_{-55}$ & 103$\pm$106 & 120 & 0.17 & 0.063 & 0.853 \\
		
		Abell 119 & 00 56 21.37 & -01 15 46 & 3.30$^{a}$ & 13190$\pm$77 & 60 & 760$^{+47}_{-58}$ & 222$\pm$92 & 220 & 0.18 & 0.091 & 0.579 \\
		
		Abell 602 & 07 53 19.02 & +01 57 25 & 1.12$^{b}$ & 18587$\pm$94 & 34 & 626$^{+55}_{-75}$ & 147$\pm$112 & 110 & 0.69 & 0.484 & 0.163 \\
		
		Abell 1066 & 10 39 23.92 & +00 20 41 & 1.20$^{c}$ & 20985$\pm$91 & 62 & 714$^{+53}_{-69}$ & 363$\pm$116 & 130 & 0.09 & 0.011 & 0.020 \\
		
		Abell 1190 & 11 11 46.22 & +02 43 23 & 1.75$^{d}$ & 22484$\pm$87 & 66 & 669$^{+51}_{-66}$ & 140$\pm$98 & 30 & 0.23 & 0.309 & 0.194 \\
		
		Abell 1205 & 11 13 22.39 & +00 10 03 & 1.77$^{c}$ & 22784$\pm$106 & 49 & 748$^{+61}_{-82}$ & 440$\pm$126 & 20 & 0.18 & 0.013 & 0.026 \\
		
		Abell 1367 & 11 44 29.53 & +01 19 21 & 1.25$^{a}$ & 6595$\pm$49 & 48 & 660$^{+31}_{-37}$ & 200$\pm$72 & 320 & 0.06 & 0.007 & 0.026 \\
        
        Abell 1589 & 12 41 35.79 & +01 14 22 & 1.53$^{e}$ & 21585$\pm$88 & 74 & 751$^{+52}_{-66}$ & 140$\pm$103 & 160 & 0.17 & 0.316 & 0.124 \\
		
		Abell 1650 & 12 58 46.20 & -01 45 11 & 6.99$^{a}$ & 25182$\pm$100 & 51 & 670$^{+57}_{-77}$ & 268$\pm$109 & 0 & 0.10 & 0.215 & 0.636 \\
		
		Abell 1656 & 12 59 48.73 & +27 58 50 & 7.77$^{a}$ & 6895$\pm$40 & 150 & 817$^{+26}_{-29}$ & 37$\pm$57 & 350 & 0.41 & 0.229 & 0.087 \\
		
		Abell 1668 & 13 03 51.41 & +01 17 04 & 1.71$^{d}$ & 18886$\pm$89 & 47 & 639$^{+52}_{-69}$ & 112$\pm$115 & 40 & 0.28 & 0.540 & 0.336 \\
		
		Abell 1773 & 13 42 08.59 & +00 08 59 & 1.37$^{c}$ & 22784$\pm$96 & 68 & 687$^{+55}_{-73}$ & 149$\pm$114 & 260 & 0.19 & 0.289 & 0.336 \\
		
		Abell 1795 & 13 49 00.52 & +26 35 06 & 10.26$^{a}$ & 18587$\pm$92 & 72 & 785$^{+55}_{-69}$ & 246$\pm$108 & 180 & 0.13 & 0.031 & 0.265 \\
		
		Abell 1809 & 13 53 06.40 & +00 20 36 & 1.69$^{e}$ & 23683$\pm$80 & 64 & 618$^{+46}_{-60}$ & 43$\pm$95 & 270 & 0.17 & 0.101 & 0.420 \\
		
		Abell 2029 & 15 10 58.70 & +05 45 42 & 17.44$^{a}$ & 23084$\pm$102 & 117 & 893$^{+60}_{-76}$ & 79$\pm$111 & 10 & 0.29 & 0.370 & 0.415 \\
		
		Abell 2052 & 15 16 45.51 & +00 28 00 & 2.52$^{a}$ & 10492$\pm$65 & 38 & 619$^{+40}_{-50}$ & 48$\pm$71 & 320 & 0.21 & 0.129 & 0.663 \\
        
		Abell 2061 & 15 21 15.31 & +30 39 16 & 4.85$^{f}$ & 23383$\pm$69 & 91 & 630$^{+41}_{-51}$ & 154$\pm$91 & 210 & 0.09 & 0.043 & 0.183 \\
		
		Abell 2063 & 15 23 01.87 & +00 34 34 & 2.19$^{a}$ & 10492$\pm$78 & 58 & 785$^{+48}_{-59}$ & 163$\pm$93 & 330 & 0.35 & 0.170 & 0.016 \\
        
        Abell 2065 & 15 22 42.60 & +27 43 21 & 5.55$^{a}$ & 21884$\pm$98 & 113 & 873$^{+58}_{-73}$ & 422$\pm$125 & 10 & 0.07 & 0.002 & 0.211 \\ 
        
        Abell 2069 & 15 23 57.94 & +01 59 34 & 3.45$^{g}$ & 34775$\pm$139 & 69 & 910$^{+77}_{-104}$ & 363$\pm$178 & 150 & 0.21 & 0.089 & 0.179 \\
        
        Abell 2107 & 15 39 47.92 & +01 27 05 & 1.41$^{e}$ & 12291$\pm$62 & 42 & 615$^{+38}_{-47}$ & 159$\pm$74 & 280 & 0.15 & 0.021 & 0.151 \\
        
        Abell 2124 & 15 44 59.33 & +02 24 15 & 1.66$^{f}$ & 19786$\pm$103 & 53 & 751$^{+60}_{-80}$ & 38$\pm$130 & 150 & 0.45 & 0.705 & 0.873 \\
		
		Abell 2199 & 16 28 38.50 & +39 33 60 & 4.09$^{a}$ & 8993$\pm$52 & 75 & 649$^{+33}_{-39}$ & 156$\pm$59 & 10 & 0.13 & 0.008 & 0.586 \\
		
        Abell 2670 & 23 54 10.15 & -00 41 37 & 2.28$^{c}$ & 22784$\pm$89 & 92 & 799$^{+53}_{-66}$ & 232$\pm$104 & 220 & 0.11 & 0.085 & 0.523 \\
        
		ZWCL1215 & 12 17 41.44 & +03 39 32 & 5.17$^{a}$ & 22484$\pm$86 & 87 & 760$^{+51}_{-64}$ & 58$\pm$118 & 240 & 0.40 & 0.888 & 0.873 \\
        
		\hline
	\end{tabular}
	\end{sideways}
	\begin{tablenotes}
		\item $^{a}$ \cite{Reiprich2002} \hspace{0.45cm} $^{e}$ \cite{Jones1999}
		\item $^{b}$ \cite{Ebeling1998} \hspace{1.4cm} $^{f}$ \cite{Marini2004}
		\item $^{c}$ \cite{Popesso2007c} \hspace{1.4cm} $^{g}$ \cite{David1999}
		\item $^{d}$ \cite{Boehringer2000}
	\end{tablenotes}
\end{table*}


\begin{table*}
    \centering
    \caption{The volume-limited NED cluster sample as per Pimbblet (2008). The descriptors for the NED cluster kinematic values can be found in Table \ref{tab:bax}.}
    \begin{tabular*}{\textwidth}{l@{\extracolsep{\fill}}c c c c c c c c c r}
    \hline
    Cluster & RA & DEC &  $\overline{cz}_{\text{glob}}$ & $N_{r_{200}}$ & $\sigma_{r_{200}}$ & $v_{\text{glob}}$ & $\theta_{\text{glob}}$ & $\chi^{2}_{id}/\chi^{2}_{rd}$ & $P(KS)$ & $P(\Delta)$ \\
    & (J2000) & (J2000) & (km s$^{-1}$) & & (km s$^{-1}$) & (km s$^{-1}$) & ($^{\circ}$) & & & \\
    \hline
    Abell 533 & 05 01 30.79 & -01 30 27 & 14000$\pm$171 & 22 & 751$^{+99}_{-153}$ & 951$\pm$319 & 270 & 0.09 & 0.002 & 0.026 \\
    Abell 2860 & 01 04 20.62 & -02 39 17 & 31718$\pm$48 & 14 & 229$^{+27}_{-39}$ & 131$\pm$60 & 210 & 12.92 & 0.035 & 0.878 \\
    Abell 2911 & 01 26 04.60 & -02 31 54 & 24223$\pm$85 & 31 & 484$^{+49}_{-68}$ & 250$\pm$118 & 50 & 0.19 & 0.058 & 0.129 \\
    Abell 3151 & 03 40 27.71 & -01 54 49 & 20265$\pm$115 & 50 & 753$^{+67}_{-91}$ & 278$\pm$168 & 10 & 0.09 & 0.227 & 0.041 \\
    Abell 3266 & 04 31 24.10 & -04 05 47 & 17657$\pm$54 & 281 & 825$^{+34}_{-38}$ & 55$\pm$81 & 200 & 0.13 & 0.219 & 0.356 \\
    Abell 3391 & 06 26 22.80 & -03 34 47 & 15409$\pm$114 & 81 & 931$^{+69}_{-88}$ & 525$\pm$151 & 240 & 0.08 & 0.003 & $\ll$0.01 \\
    Abell 3528 & 12 54 18.20 & -01 56 05 & 15829$\pm$71 & 103 & 674$^{+43}_{-53}$ & 48$\pm$89 & 140 & 3.61 & 0.663 & 0.834 \\
    Abell 3570 & 13 46 52.50 & -02 31 29 & 10972$\pm$50 & 16 & 233$^{+30}_{-45}$ & 110$\pm$81 & 150 & 1.09 & 0.023 & 0.612 \\
    Abell 3535 & 12 57 48.55 & -01 53 57 & 19546$\pm$53 & 28 & 291$^{+31}_{-43}$ & 87$\pm$69 & 30 & 1.15 & 0.402 & 0.319 \\
    Abell 3653 & 19 53 00.90 & -03 28 07 & 32647$\pm$84 & 43 & 565$^{+48}_{-63}$ & 261$\pm$114 & 170 & 0.10 & 0.094 & 0.115 \\ 
    Abell 3716 & 20 51 16.70 & -03 30 47 & 13850$\pm$76 & 117 & 767$^{+47}_{-57}$ & 200$\pm$110 & 220 & 0.06 & 0.152 & 0.586 \\
    Abell 3744 & 21 07 12.29 & -01 41 45 & 11422$\pm$60 & 64 & 481$^{+37}_{-47}$ & 61$\pm$76 & 320 & 0.17 & 0.928 & 0.028 \\
    \hline
    \end{tabular*}
    \label{tab:ned}
\end{table*}

\section{Cluster Rotation \& Environmental Effects}
\label{sec:cree}

It has already been noted in section \ref{sec:intro} how one might expect galaxy clusters to have gained their momentum.
To study how different cluster activity states drive the dynamical side-effects onto cluster galaxy sub-populations we utilise the BAX-selected clusters to compile two sub-samples, those that possess sub-structure against those without, as determined by the \cite{Dressler1988} test ($\Delta$-test hereafter) which is outlined in section \ref{ssec:ds-test} below.
The methodology outlined by MP17, elaborated in section \ref{ssec:MP}, is applied to construct the rotational curves determined through their artificial transverse rotation on the plane of the sky to study how the cluster galaxy sub-populations respond to global cluster rotation and level of sub-structuring.
From this method, we determine the rotational profiles as a function of projected radius of the clusters from both our samples, as well as producing composite profiles for merging and non-merging (i.e. level of activity) sub-samples for the BAX-selected clusters (see sections \ref{ssec:dbbll} and \ref{ssec:BAX}).
Allowing for us to compare the angular momentum of galaxy clusters between different possible states of merger activity and how cluster galaxy sub-populations drive the resultant profiles.

\subsection{The $\Delta$-Test}
\label{ssec:ds-test}
The use of a geometric perspective motion effect can lead to spurious detections of ambient cluster rotation in the presence of strong galaxy-galaxy merger activity from a merging cluster. 
However, within this work, we aim to establish the rotational velocities (see \ref{ssec:MP}) as a function of radius to determine how the strength of the global cluster rotation varies between merging and non-merging environments.
Therefore, to test how cluster activity can alter the rotational dynamics of clusters and thereby, affect the evolution of the cluster galaxies, we incorporate the $\Delta$-test for sub-structure on galaxies within 1.5 Mpc $h^{-1}$ (defined as $N_{\text{glob}}$) of the BAX and NED defined cluster centres in order to delineate between merging and non-merging environments.
The $\Delta$-test is a commonly used and very robust tool for indicating substructure, with sub-structure detections reaching $>99\%$ confidence when applied on cluster galaxy sample sizes of $N_{\text{glob}}\geq60$ (e.g. \citealt{Pinkney1996,Pimbblet2008,Song2018,Bilton2018}). 
Therefore, we apply the $\Delta$-test onto each galaxy and their $N_{nn} = \sqrt{N_{\text{glob}}}$ nearest neighbours. 
The localised kinematics are determined and then compared against the global values,

\begin{equation}
    \delta^{2}_{i} = \left(\frac{N_{nn}+1}{\sigma_{\text{glob}}^{2}}\right) [(\overline{cz}_{\text{local}} - \overline{cz}_{\text{glob}})^{2} + (\sigma_{\text{local}} - \sigma_{\text{glob}})^{2}],
	\label{eq:ds}
\end{equation}

\noindent
where $\delta$ measures the deviation in the small region around the galaxy compared to the global cluster values at $\leq1.5$ Mpc $h^{-1}$.
The application of the $\Delta$-test to the BAX sample leads to a merging sub-sample size of 8 clusters and a non-merging sub-sample size of 25 clusters, the resultant values for which can be found in Table \ref{tab:bax}.
Example phase-space diagrams, along with their respective caustics, produced from each sub-sample of clusters are presented in Figure \ref{fig:caustics} to illustrate the membership and spread of galaxies for each cluster and their local environments.

\subsection{The Manolopoulou \& Plionis Method}
\label{ssec:MP}
In order to determine our averaged cluster galaxy rotational profiles, we employ the methodology of MP17, which utilises the geometrical \textquoteleft Perspective Rotation'.
Assuming an ideal case where the rotational axis of a cluster is perpendicular to our line-of-sight (i.e. $\phi=0^{\circ}$), we split our cluster into two semicircles vertically down the X-ray defined centre and determine their line-of-sight velocities of the member galaxies with respect to their angle $\mu$ from the origin.
The mean velocity of each semicircle ($\langle v_{1} \rangle,\langle v_{2} \rangle$) is then determined in equation \ref{eq:vels}. 
Enabling observations in how the difference in the mean velocities of each semicircle ($v_{diff} = \langle v_{1} \rangle-\langle v_{2} \rangle$) vary as we project the average proper motions of the galaxies through the transverse rotation of galaxies in $\theta=10^{\circ}$ increments.
Therefore, for each semicircle we apply iteratively 

\begin{equation}
 	\langle v_{1,2} \rangle = \frac{1}{N} \sum_{i=1}^{N} \Delta V_{i} \cos(90^{\circ}-\mu_{i}),
    \label{eq:vels}
\end{equation}

\noindent
where $\Delta V_{i}$ is the line-of-sight velocity from equation \ref{eq:dv} for the galaxy $z_{\text{gal,}i}$, and the angle from the origin $\mu_{i}$ operates between $0^{\circ}$ and $180^{\circ}$ for each semicircle.
This means that $v_{\text{diff}}$ can be determined for each angle $\theta$. 
Leading to the uncertainties of each semicircle being propagated through for each angle $\theta$ as

\begin{equation}
    \sigma_{\theta} = \sqrt{\frac{\sigma^{2}_{v,1}}{n_{1}} + \frac{\sigma^{2}_{v,2}}{n_{2}}},
    \label{eq:unc}
\end{equation}

\noindent
where $\sigma_{v}$ is the velocity dispersion and $n$ is the galaxy number for each semicircle 1 and 2 at each angle of $\theta$.


\begin{figure*}
    \centering
    \includegraphics[width=0.85\paperwidth]{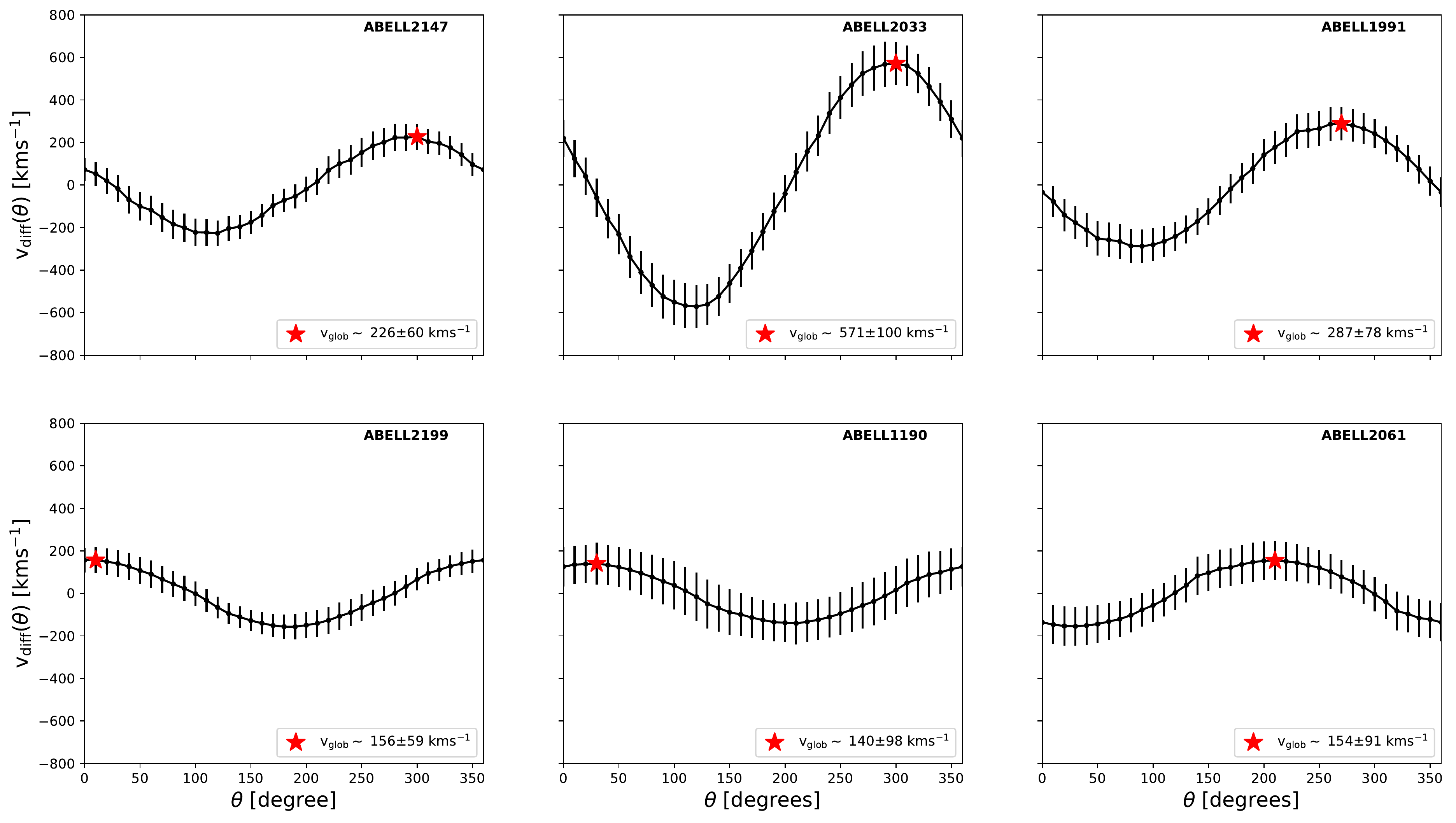}
    \caption{A selection of example BAX cluster sinusoidal rotational curves of merging (top row) and non-merging (bottom row) clusters, as determined by the $\Delta$-test for sub-structure with galaxies that lie $\leq1.5$ Mpc $h^{-1}$ from the cluster centre, with $v_{\text{diff}}$ as a function of $\theta$ as per the MP17 methodology outlined in section \ref{ssec:MP}. The red star marks the point at which $v_{\text{glob}} =$ MAX$[v_{\text{diff}}(\theta)]$. The uncertainties on the real data curve are derived by the propagation of the standard error as denoted in Equation \ref{eq:unc}.}
    \label{fig:examvdif}
\end{figure*}

Finally, we assume the maximum $v_{\text{diff}}(\theta)$ provides the rotational velocity $v_{\text{rot}} =$ MAX$[v_{\text{diff}}(\theta)]$, which consequently, provides the angle of the rotational axis in the plane of the sky $\theta_{\text{rot}}$.
Therefore, for our global cluster definition, we determine rotational values for each cluster from our BAX and NED samples that are computed using equation \ref{eq:vels} with galaxies that lie $\leq1.5$ Mpc $h^{-1}$ from their respective cluster centres.
Thus, providing the final cumulative global cluster rotational velocities and angle of the rotational axes for each, which are denoted as $v_{\text{glob}}$ and $\theta_{\text{glob}}$ respectively.
The statistical significance of the presence of rotation from our global definition is calculated for galaxies from both our BAX and NED cluster samples.
Following the methodologies of MP17, we determine the ideal ($\chi^{2}_{id}$) and random ($\chi^{2}_{rd}$) $\chi^{2}$ statistic as a by-product of our analysis and can be found within Table \ref{tab:bax}.
Figure \ref{fig:examvdif} presents an example of the MP17 methodology with our BAX merging and non-merging cluster sub-samples (consistent with the examples in Figure \ref{fig:caustics}) to determine cluster rotation in the form of the sinusoidal curves produced by artificially rotating the clusters in the plane of the sky.

%

\begin{figure*}
    \centering
    \includegraphics[width=0.85\paperwidth]{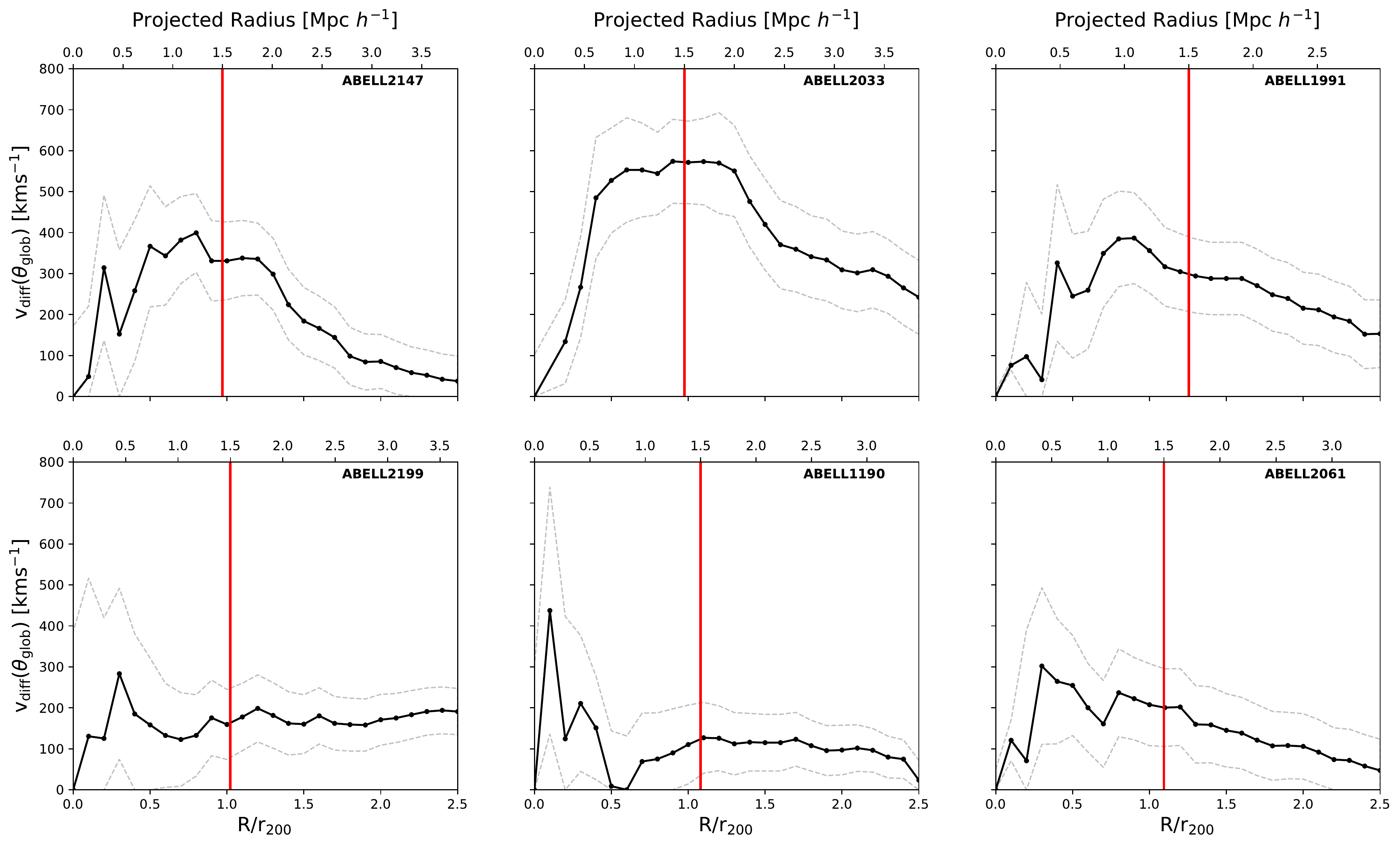}
    \caption{A selection of example $v_{\text{diff}}(\theta_{\text{glob}})$ rotational profiles consistent with Figure \ref{fig:examvdif}, as a function of the projected virial radius $R/r_{200}$ in increments of $0.1$, of merging (top row) and non-merging (bottom row) as determined by the $\Delta$-test for sub-structure. The red vertical lines represent the point at which the global cluster values are determined and the statistical tests for both sub-structure and rotation are calculated with galaxies that $\leq1.5$ Mpc $h^{-1}$ from the cluster centre. Note the consistency of higher rotational velocity throughout the merging clusters in comparison to the dampened profiles for the non-merging clusters. The dashed lines represent the uncertainties derived from the propagated standard error as denoted in Equation \ref{eq:unc}.}
    \label{fig:examvrot}
\end{figure*}

The thesis presented here is focused on how $v_{\text{diff}}$ is dependent on the cluster galaxy sub-populations as a function of cluster radius at different epochs; core-merging events between two originally independent clusters; sub-structuring of galaxies relaxing from a core-merger event; older and relaxed clusters that are homogeneous to our tests of sub-structure.
We therefore, using the calculated global cluster defined values, determine how $v_{\text{glob}}$ varies as a function of radius from the cluster centre in incremental units of $0.1 r_{200}$ with a coverage of $0<R\leq 2.5 r_{200}$ by fixing our theta to the rotational axis $\theta_{\text{glob}}$.
An example of the application of this methodology to the individual BAX-defined clusters between merging and non-merging environments, as defined previously by the $\Delta$-test, is depicted in Figure \ref{fig:examvrot}. 

Here we find the merging clusters demonstrate rising profiles from the cluster centre that lead to consistently high $v_{\text{diff}}(\theta_{\text{glob}})$ values throughout to $2.5 r_{200}$.
In contrast, non-merging clusters possess dampened core-rotational velocities, with Abell 2199 showing a consistent profile out towards $2.5 r_{200}$, most likely as a result of the outer galaxy members homogenising with the cluster's angular momentum. 
The behaviour observed here between the two sub-samples runs parallel to the $\Delta$-test for sub-structure; increasing core-rotational velocities show correlation with merging environments. 
This response of environment to the rotational velocities is not completely surprising considering both methodologies are constrained to the same projected radii and radial velocity measurements.
It should be noted that this effect is not completely consistent to every $\Delta$-test defined merging cluster, which highlights the limitations of analysing 3D motions through a projected 2D-plane of sky.

\subsection{Dumbbell BCG Clusters}
\label{ssec:dbbll}
\begin{figure}
    \centering
    \includegraphics[width=0.85\columnwidth]{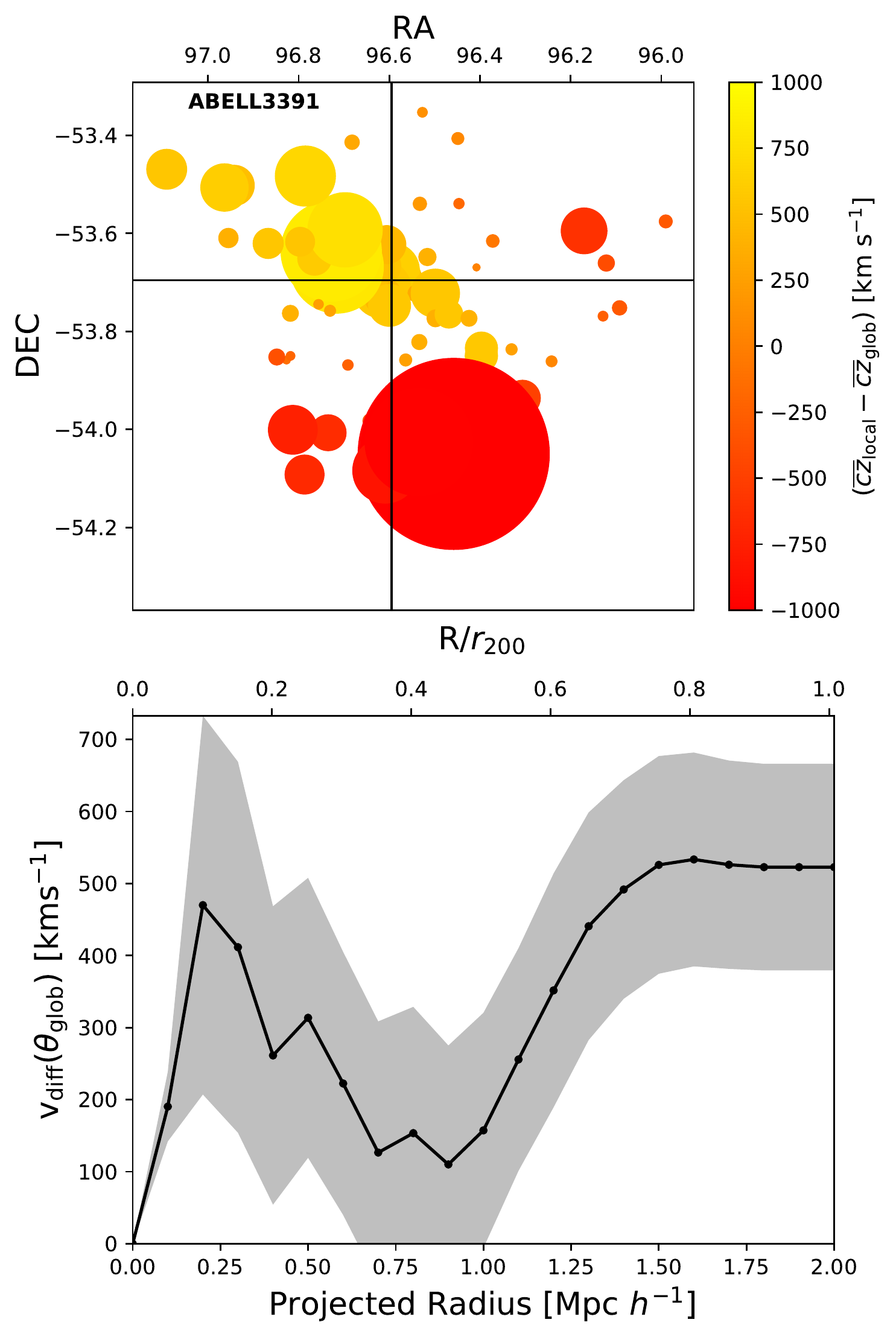}
    \caption{Top: bubble plot of Abell 3391 from the $\Delta$-test, where the size of each circle is proportional to $\pi (e^{\delta_{i}})^{2}$, the black cross represents the NED-defined centre and the colours representing varying radial velocity differences $[\overline{cz}_{\text{local}} - \overline{cz}_{\text{glob}}]$. Bottom: the rotational profile of Abell 3391.}
    \label{fig:3391}

\end{figure}

\begin{figure}
    \centering
    \includegraphics[width=0.85\columnwidth]{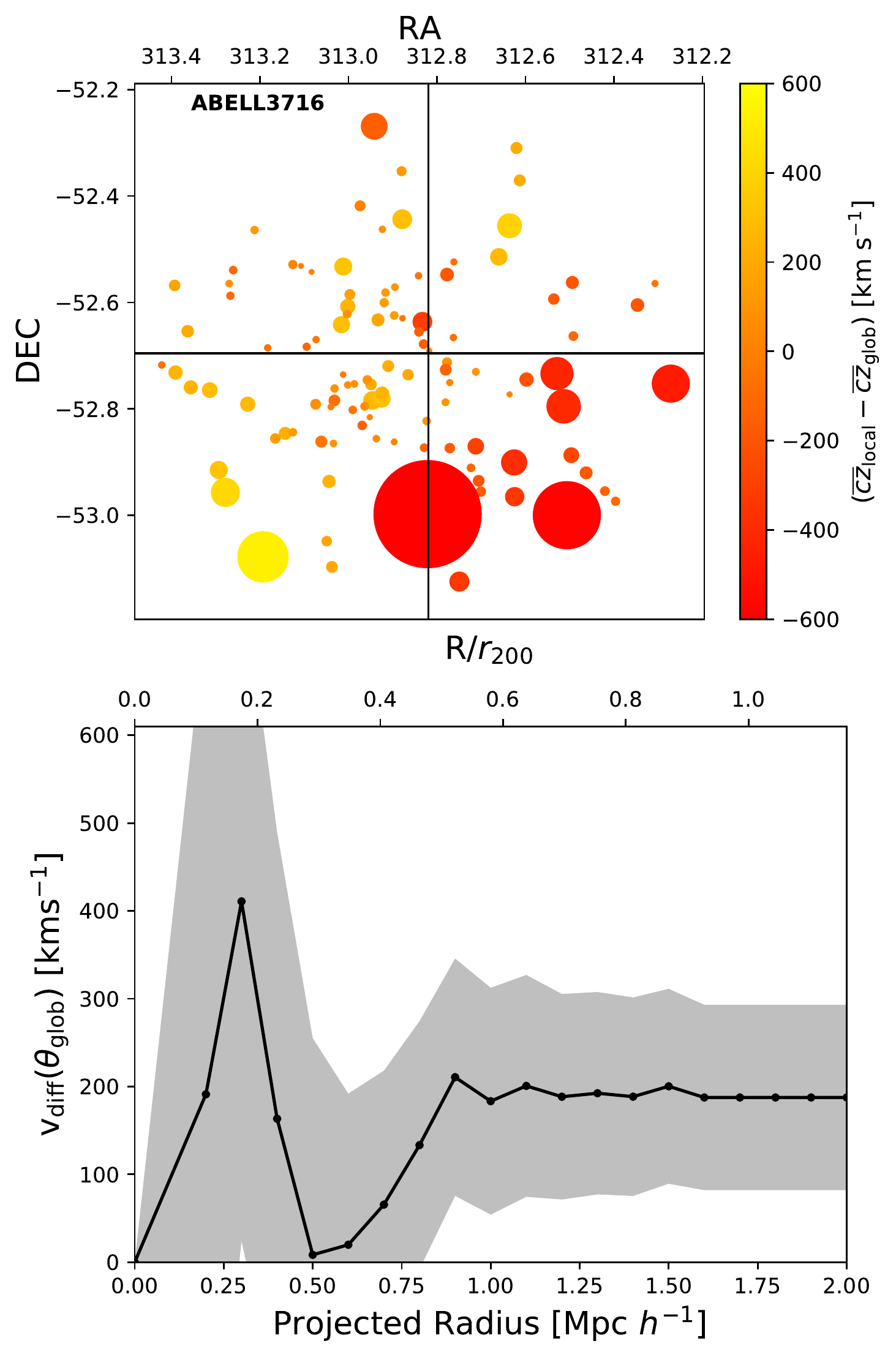}
    \caption{Top: bubble plot of Abell 3716 from the $\Delta$-test, where the size of each circle is proportional to $\pi (e^{\delta_{i}})^{2}$, the black cross represents the NED-defined centre and the colours representing varying radial velocity differences $[\overline{cz}_{\text{local}} - \overline{cz}_{\text{glob}}]$. Bottom: the rotational profile of Abell 3716.}
    \label{fig:3716}
\end{figure}

\begin{figure}
    \centering
    \includegraphics[width=0.85\columnwidth]{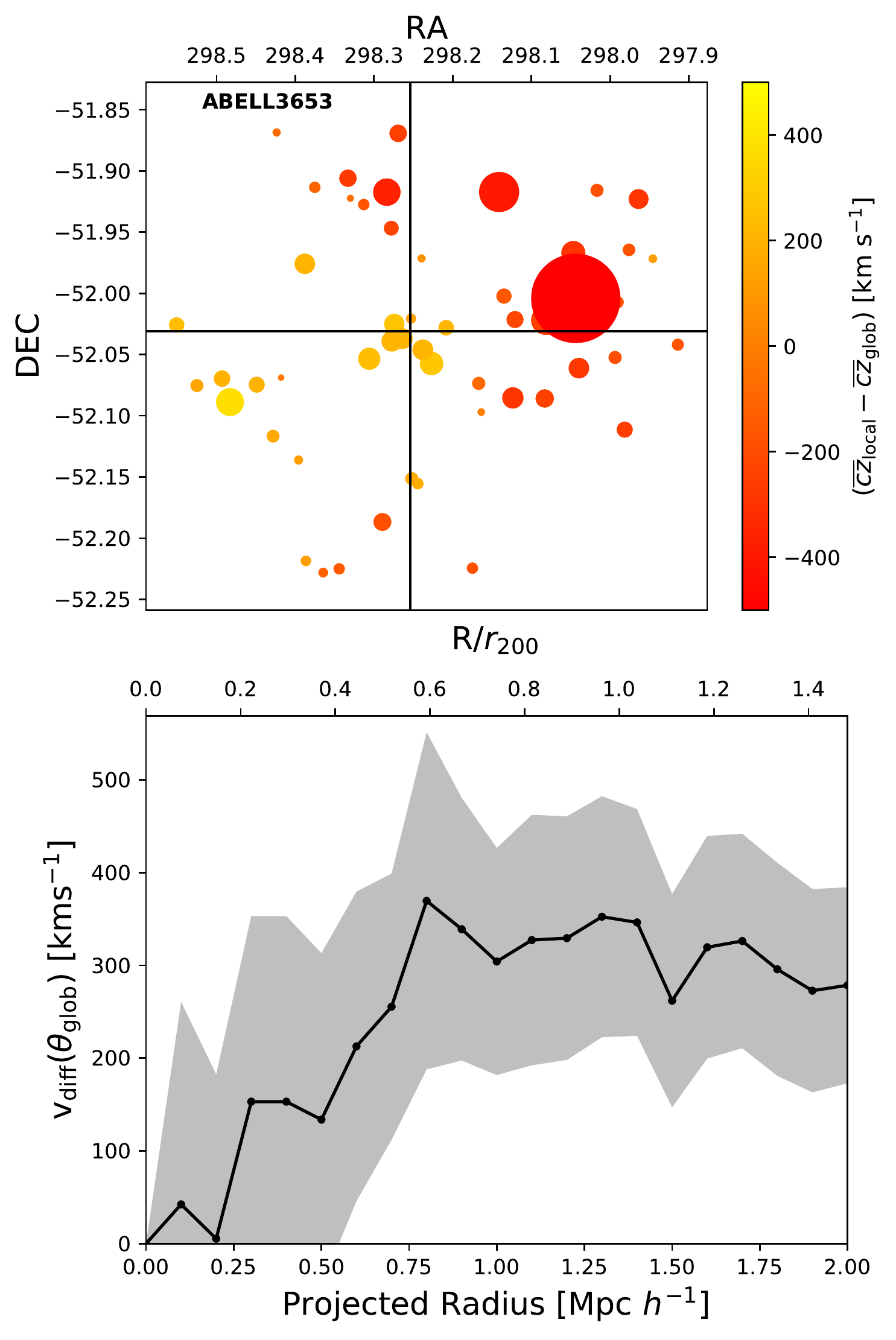}
    \caption{Top: bubble plot of Abell 3653 from the $\Delta$-test, where the size of each circle is proportional to $\pi (e^{\delta_{i}})^{2}$, the black cross represents the NED-defined centre and the colours representing varying radial velocity differences $[\overline{cz}_{\text{local}} - \overline{cz}_{\text{glob}}]$. Bottom: the rotational profile of Abell 3653.}
    \label{fig:3653}
\end{figure}

If we assume that the evolution of angular momentum within clusters originates from the off-axial interaction between two smaller clusters, then this could potentially be detected through line-of-sight measurements that are sensitive to determining rotation.
Hence, we consider clusters that host multiple BCG components with significant velocity offsets could be the result of recently merged sub-clusters that are relaxing onto a common potential.
Therefore, we elected to study a sample of dumbbell BCG clusters for their global rotational profiles with the aim to test if their offset peculiar velocity BCG cores are an indicator of higher levels of merger activity; resembling earlier epochs of post-merger relaxation.
Using the volume-limited sample as outlined in section \ref{sec:data} with the NED galaxies we perform the $\Delta$-test for substructure from the NED centres to $\leq1.5$ Mpc $h^{-1}$ (see \ref{ssec:ds-test}). 
A comparison between the statistical results, alongside the bubble plots, and the global rotational profiles, as determined in section \ref{ssec:MP}, of each dumbbell BCG cluster is made.
An example of these results for our dumbbell BCG hosting clusters can be found in Figures \ref{fig:3391}, \ref{fig:3716} and \ref{fig:3653}. 



We, unsurprisingly, do not find a significant correlation between sub-structuring and the presence of multiple off-set velocity BCGs, consistent with the findings of \cite{Pimbblet2008}.
This is in despite of the use of a standard, more loose, criteria where in this work sub-structure is deemed significant at $P(\Delta)\leq0.01$.
It is more likely that sub-structure would not play a key role in the instance of dumbbell cluster BCG cores due to the collisionless nature of galaxies on the timescales presented, especially if the dumbbell cores are in the early stages of a merger between two initially independent potential wells.
Abell 3391 is the only cluster in the dumbbell sample found to possess sub-structure, the bubble plot is presented, along with the rotational profile, in Figure \ref{fig:3391}.
It is interesting to note how the rotational profile of Abell 3391 decreases to a minimum at $\sim1.0$ Mpc $h^{-1}$, before rising back to previous $v_{\text{glob}}$ values within the same projected radial separation.
This, compared with the substantial sub-structuring observed with the bubble plot, illustrates a strong double-component system of rotating galaxies; a fast rotating core and a fast rotating outer region as a result of an on-going active merging event between two originally independent BCGs and their host galaxy clusters.
This inference is exacerbated by comparing the same analysis the dumbbell hosting clusters of Abell 3716 and Abell 3653, which can be seen in Figures \ref{fig:3716} and \ref{fig:3653} respectively.
The $v_{\text{glob}}$ profile of Abell 3716 in particular marries closely to that of Abell 3391, decreasing to a minimum $\lesssim 1.0$ Mpc $h^{-1}$, before only subtly rising back to dampened levels of rotation where $v_{\text{glob}}\sim200$kms$^{-1}$.
From the bubble plot of Abell 3716 we can see there are some small pockets of deviation from the global values, although, not to the levels found in Abell 3391.
We can, therefore, surmise that the Abell 3391 and 3716 are dumbbell hosting clusters in different stages of merging, where the Abell 3391 is in an active phase of merging with intense galaxy-galaxy interactions providing off-axial angular momentum donation. With Abell 3716 in an earlier, less-active phase of merging, where the galaxies are still yet to interact due to their collisionless nature.
Abell 3653 presents a $v_{\text{glob}}$ profile with a remarkably consistent zero gradient with exception of the bulk increase in rotation at $0.5\lesssim R \lesssim 1.0$ Mpc $h^{-1}$.
It is also notable how the bubble plot of Abell 3653 has overt displays of sub-structure towards the west of the sky, where aside from the sizeable peculiar velocity of the BCG addressed in the study of \cite{Pimbblet2006}, X-ray analysis conducted by \cite{Caglar2017} has shown the location of this sub-structure to be coincident with another BCG hosting sub-cluster.
As further stated within the work of \cite{Caglar2017} the presence of harder X-ray emission in the space in-between these two independent BCGs, along with their $\sim35$kpc off-set from their respective X-ray peaks, is a shock region between their ICM environments indicative of an on-going initial merger phase.
Taking into account the projected radial separation between the two BCGs of these sub-clusters ($\sim500$kpc), we can see that the bulk rotation observed $0.5 \lesssim R \lesssim 0.8$Mpc $h^{-1}$ is primarily the result of the foreground BCG towards the west of the sky combined with host sub-cluster's members, as can be seen by the apparent dichotomy in the radial velocities.
The clear background and foreground structures, along with the even delineation between them, throughout the projected radius studied creates the impression of a consistent $v_{\text{glob}}$ profile, aside from the boost provided by sub-structured west BCG $\sim0.5$Mpc h$^{-1}$, which introduces the possibility that the global rotation via our method is merely the result of the z-space difference between the two sub-clusters.
With this in mind, however, \cite{Caglar2017} have concluded the sub-clusters are gravitationally bound and are in infall at 2400 km s$^{-1}$ with core passage expected in 380 Myr.
The peculiar velocities of the dumbbell components of each cluster shown here, along with the two independent Abell 3653 BCG components, are detailed in Table \ref{tab:pecv} for easy comparison.
Considering the results from the examples shown here the varying rotation profile, close angular separation between the cores and levels of sub-structuring present in Abell 3391, the cluster must be in a \textquoteleft post-initial merger' phase; the two cores are relaxing onto a common potential with the surrounding population of galaxies aggressively interacting with one another as a result of their latent friction and global rotation donated from the initial merging phase.


\begin{table}
    \centering
    \caption{The peculiar velocities of the example dumbbell hosting clusters presented in Figures \ref{fig:3391}, \ref{fig:3716} and \ref{fig:3653} within the reference frames of their respective $\overline{cz}$ values, referenced from Pimbblet (2008). The literary values for the BCGs of Abell 3653 are utilised from Caglar \& Hudaverdi (2017).}
    \begin{tabular}{lccl}
         \hline
         Cluster & RA & DEC & |$\Delta cz$| \\
         components & (J2000) & (J2000) & (km s$^{-1}$) \\
         \hline
         Abell 3391 DBL1 & 06 26 20.22 & -53 14 57.84 & $489\pm133$ \\
         \hspace{1.4cm} DBL2 & 06 26 17.80 & -53 14 56.04 & $68\pm142$ \\
         Abell 3653 DBL1 & 19 53 03.48 & -52 07 58.80 & $736\pm105$ \\
         \hspace{1.4cm} DBL2 & 19 53 02.76 & -52 08 06.00 & $495\pm126$ \\
         \hspace{1.4cm} BCG1 & 19 53 01.90 & -52 59 13.00 & $683\pm96$ \\
         \hspace{1.4cm} BCG2 & 19 52 17.30 & -51 59 50.00 & $43\pm124^{*}$ \\
         Abell 3716 DBL1 & 20 52 00.48 & -52 16 18.48 & $559\pm92$ \\
         \hspace{1.4cm} DBL2 & 20 51 66.88 & -52 16 15.60 & $255\pm88$ \\
         \hline
    \end{tabular}
    \begin{tablenotes}
        \item $^{*}$ The uncertainty is propagated through from the literary values.
    \end{tablenotes}
    \label{tab:pecv}
\end{table}

\subsection{BAX Cluster Stacks}
\label{ssec:BAX}

In order to test the dynamical evolution of clusters more generally, we make attempts to observe any contrast in the global rotation profiles across differing cluster environments that could represent different epochs of cluster-cluster merging. Therefore, to build this general picture, we build composite clusters from the BAX sample between those defined as either merging or non-merging, which has the primary benefit of boosting the signal-to-noise for the rotational profiles.
For the purposes of calculating $v_{\text{rot}}$ using MP17, following the outlined procedure in section \ref{ssec:MP}, we initiate the following stacking procedure: each cluster is rotated by their respective $\theta_{\text{glob}}$ so the rotation axis of each cluster overlaps, we then stack our clusters onto a common RA-DEC grid normalised to each cluster's respective BAX centres along with their normalised $\Delta$V values as per equation \ref{eq:vels}.
This will lead to the rotational axis of each composite stack becoming $\sim 0^{\circ}$, which provides, $v_{\text{rot}}= v_{\text{diff}}(\theta=0)$.
The final galaxy contributions to the merging and non-merging stacks are 1286 and 3349 galaxies respectively.

\begin{figure*}
    \centering
    \includegraphics[width=0.85\paperwidth]{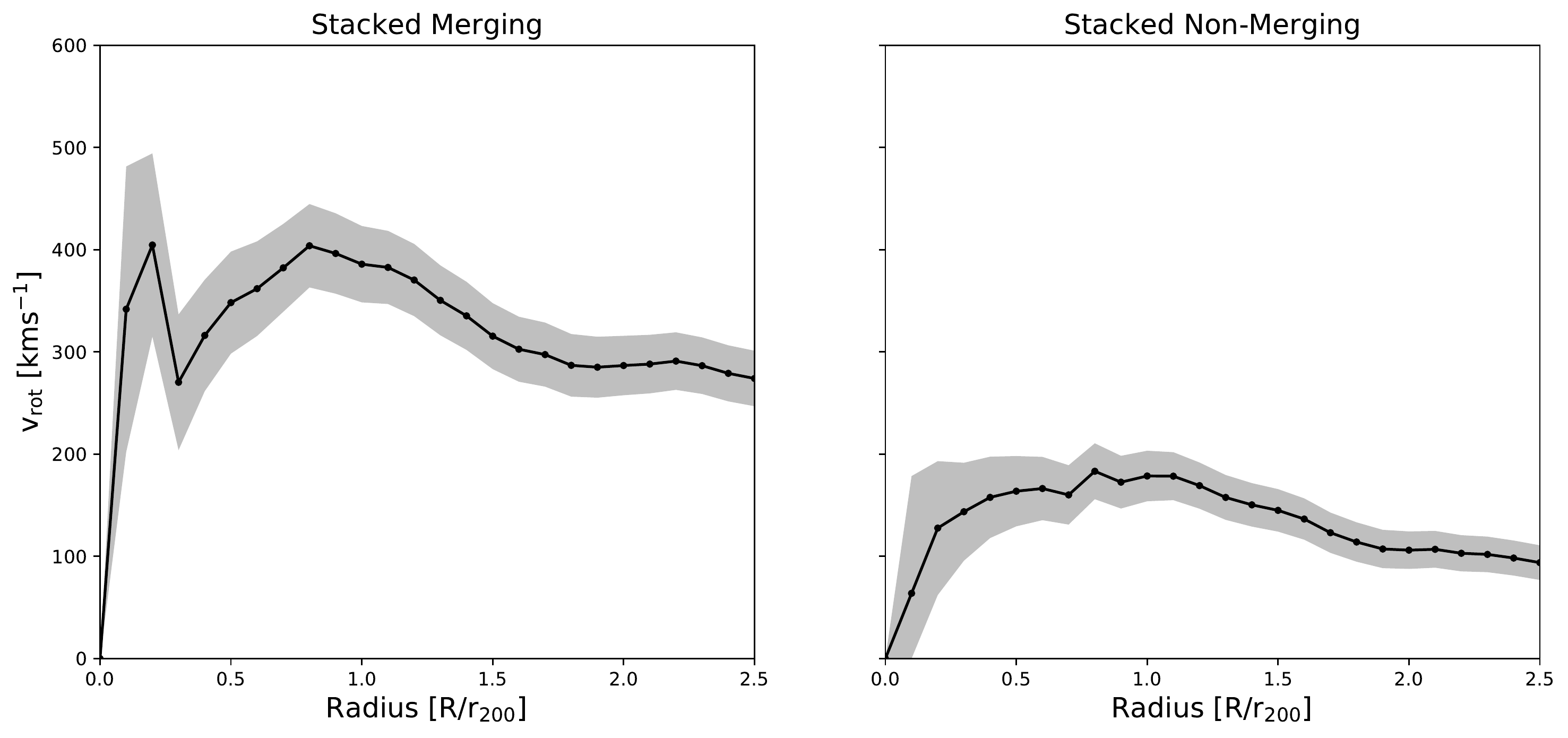}
    \caption{Composite rotational velocity profiles of all galaxies as a function of radius ($R/r_{200}$). The merging rotational profile (left) displays a high rotational velocity at the core $\lesssim 0.5 R/r_{200}$ . In contrast the non-merging rotational profile (right) shows dampened core rotation, which collapses close to zero at radii $\geq 0.5 R/r_{200}$. The shaded regions represent the uncertainties derived from the propagated standard error as denoted in Equation \ref{eq:unc}.}
    \label{fig:cvrot}
\end{figure*}

The first result of the complete merging and non-merging composites are highlighted for comparison in Figure \ref{fig:cvrot}.
We can immediately see that the rotational profile of the merging composite in Figure \ref{fig:cvrot} possesses very high core rotation peaking up to $\sim400$kms$^{-1}$ within $\lesssim 0.5 R/r{200}$.
This result is indicative of the merging sub-sample primarily consisting of relaxing galaxy cluster cores that have undergone recent core-merging processes.
The continued high $v_{\text{rot}}$ retained throughout to $\lesssim R/r_{200}$ implies this angular momentum donation mechanism is dominant.
Where the gradual decline in $v_{\text{rot}}$ at $\gtrsim R/r_{200}$ is the result of a decrease in cluster galaxy density, and therefore, interaction probability between them.
Contrarily the rotational profile of the non-merging composite presents a dampened core rotation.
The immediate inference of this dampening effect presents the non-merging composite to mainly consist of older, more evolved, clusters in more advanced stages of relaxation processes.
However, there is the need to consider that the differences in our observations of $v_{rot}$ between the merging and non-merging galaxies are down to their potential difference in mass distributions. 

To see how cluster galaxy colour sub-populations respond to the dynamics, activity and environment, the galaxies for the merging and non-merging composites are split into two sub-populations of colour, blue galaxies and red galaxies.
In order to account for the biasing of colour distributions with increasing galaxy log stellar mass we find a line of delineation that determines a galaxy's colour, which is computed with a $(u-r)$ colour gradient as a function of the log stellar mass.
Following the methodology of \cite{Jin2014}, the residual galaxies from the bi-modal $(u-r)$ distribution in bins of increasing stellar mass are used to output the k-corrected linear relation $(u-r)_{z=0} = 0.40[\log_{10}(M_{\ast})]-1.74$, this is further detailed in equation 4 of \cite{Bilton2018}.

\begin{figure*}
    \centering
    \includegraphics[width=0.85\paperwidth]{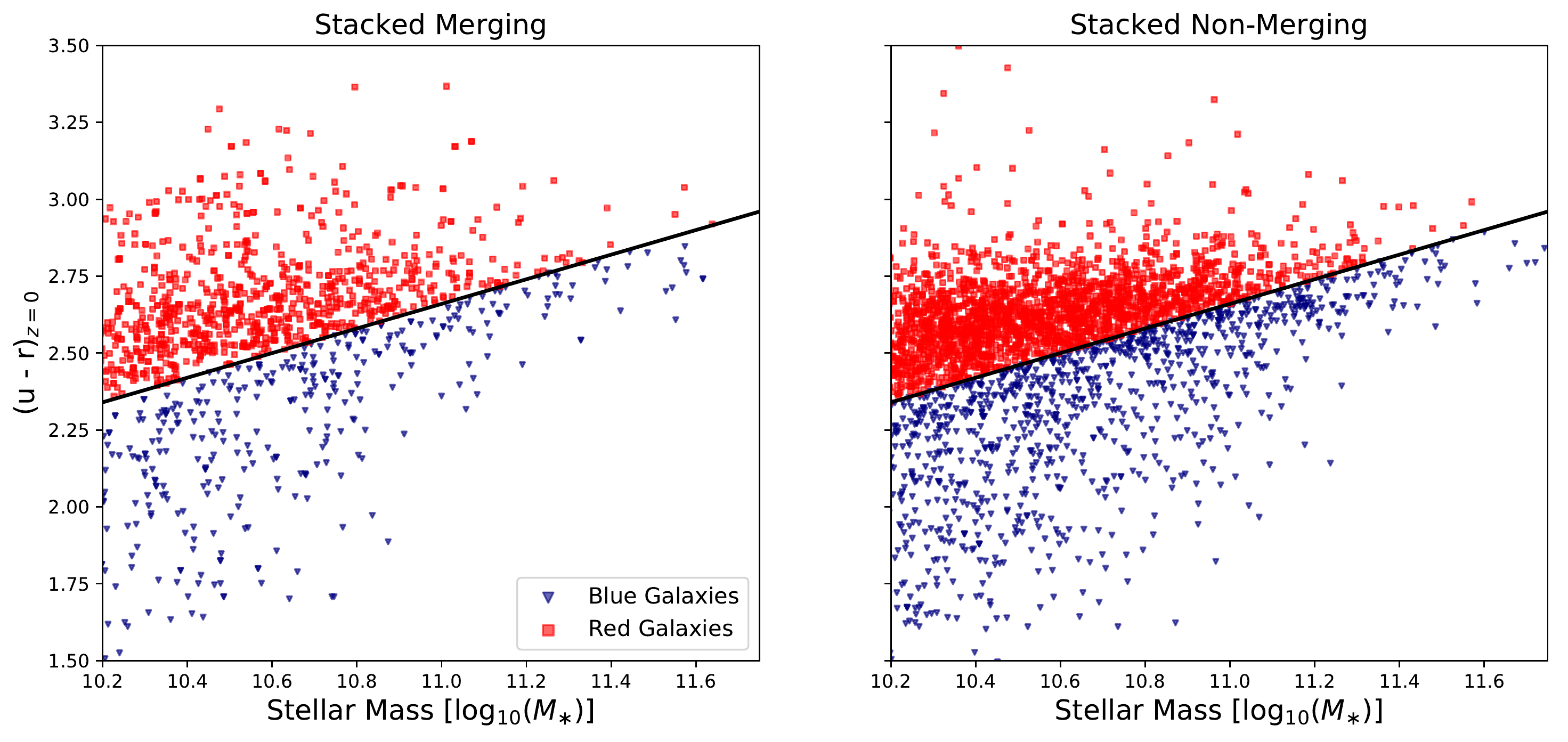}
    \caption{The colour distributions  of the mass-complete DR8 galaxies between the merging and non-merging samples: (u-r)$_{z=0}$ plotted as a function of log$_{10}$($M_{\ast}$). The black line resembles the linear fit of the centre of the bimodal distribution at quartile increments of log$_{10}$($M_{\ast}$); red galaxies are above the fitted line denoted as red squares; blue galaxies are below the fitted line denoted as blue triangles.}
    \label{fig:cldist}
\end{figure*}

\begin{figure*}
    \centering
    \includegraphics[width=0.85\paperwidth]{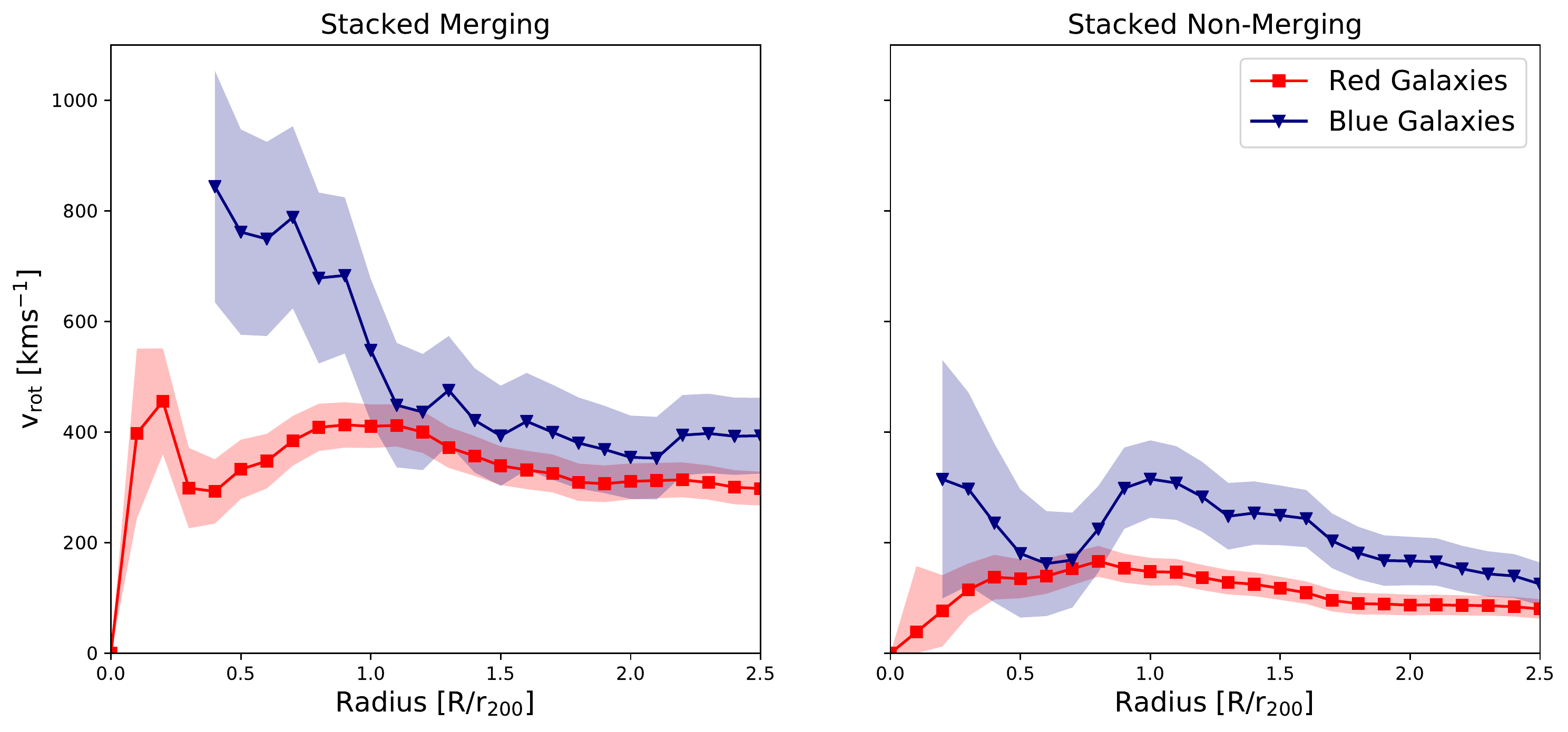}
    \caption{Composite rotational velocity profiles split by their colour with the same axes as Figure \ref{fig:cvrot}. The blue triangle and red square markers of each stack represent the blue and red galaxies respectively. 
    The blue galaxies in the merging cluster stack (left) have a high $v_{\text{glob}}$ segregation from the red galaxies at $0.4 \lesssim R/r_{200} \lesssim 1.0 r_{200}$ before homogenising $r_{200}$. The non-merging cluster stack (right) observes dampened behaviours with \textquoteleft bumps' $\gtrsim r_{200}$. The shaded regions represent the uncertainties derived from the propagated standard error as denoted in Equation \ref{eq:unc}.}
    \label{fig:cvrotcl}
\end{figure*}

All BAX clusters with their member galaxies holding DR8 photometry are k-corrected to the local rest frame $(z=0)$ before being parsed through the linear relation denoted above.
Galaxies that lie at greater values from the linear gradient are all classified as red sequence galaxies, with galaxies below classified as blue cloud galaxies.
An example of the colour distributions of the galaxies for each stack can be found in Figure \ref{fig:cldist}, providing the merging sample with 402 blue galaxies and 862 red galaxies alongside the non-merging sample with 1153 blue galaxies and 2184 red galaxies for each stack.
The $v_{\text{rot}}$ profiles are then calculated for each sub-population, and environment, utilising the same stacking and outputting sequence highlighted previously in the production of Figure \ref{fig:cvrot}.
Figure \ref{fig:cvrotcl} presents the resultant $v_{\text{rot}}$ profile with the above implemented methodology.
The merging rotational profile in Figure \ref{fig:cvrotcl} depicts the blue sub-population of galaxies with very high segregation of $v_{\text{rot}}$ values at $\lesssim R/r{200}$.
However, due to a depletion of blue galaxies towards the core there are no $v_{\text{rot}}$ values for the blue sub-population $\lesssim 0.3 R/r_{200}$.
Consequentially, this implies the observed core rotation from the merging composite in Figure \ref{fig:cvrot} is dominated by red sequence galaxies.
The immediate conclusion as a result implies that cores of merging clusters consist of evolved, red galaxies in the process of relaxing onto a new common cluster potential via \textquoteleft back and forth' sloshing motions.
The non-merging profile in Figure \ref{fig:cvrotcl} demonstrates a tighter velocity separation between the blue and red galaxy sub-populations, however, there is still a clear segregation in rotational velocity that leads to the connotations of infalling blue galaxies.
There is the significant \textquoteleft bump' in the blue galaxy sub-population at $\sim R/r_{200}$, which could inconclusively be the result of a mixture of infaller and so called \textquoteleft backsplash' galaxies within the stack (see \citealt{Pimbblet2011}).
The key result from the $v_{\text{rot}}$ composite profiles of the colour sub-populations is that the two environments are indicative of differing epochs of cluster merging; relaxed galaxies with a population of infalling, or potentially backsplash, blue-galaxies within the non-merging composite and a actively relaxing galaxies onto a common potential from successive merging processes depicting the merging composite.

\begin{figure}
    \centering
    \includegraphics[width=\columnwidth]{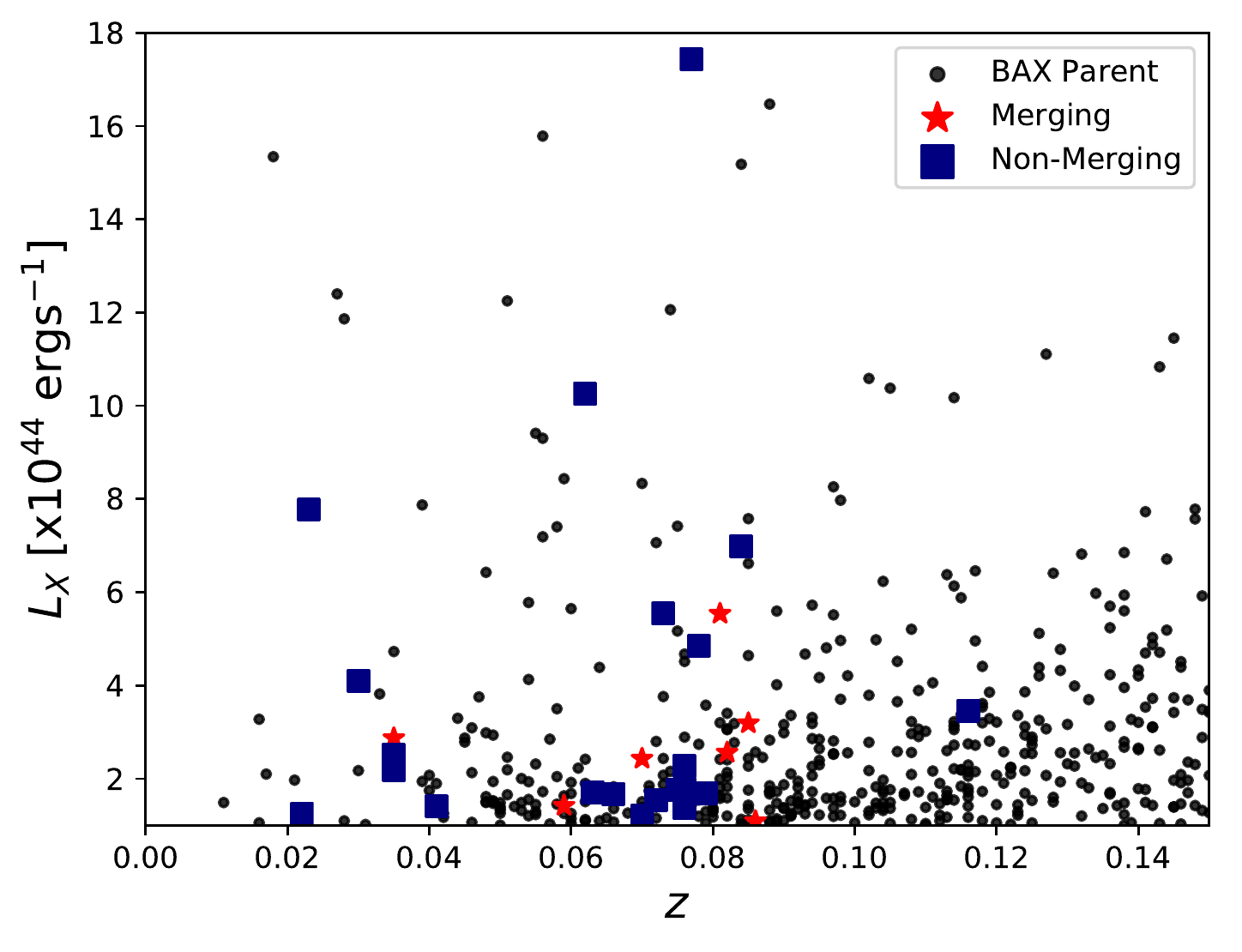}
    \caption{The distribution of BAX cluster X-ray luminosities ($L_{X}$) against redshift ($z$). Where the red stars resemble the merging sample and the blue squares depict the non-merging sample.}
    \label{fig:lxzspl}
\end{figure}

\begin{figure*}
    \centering
    \includegraphics[width=0.85\paperwidth]{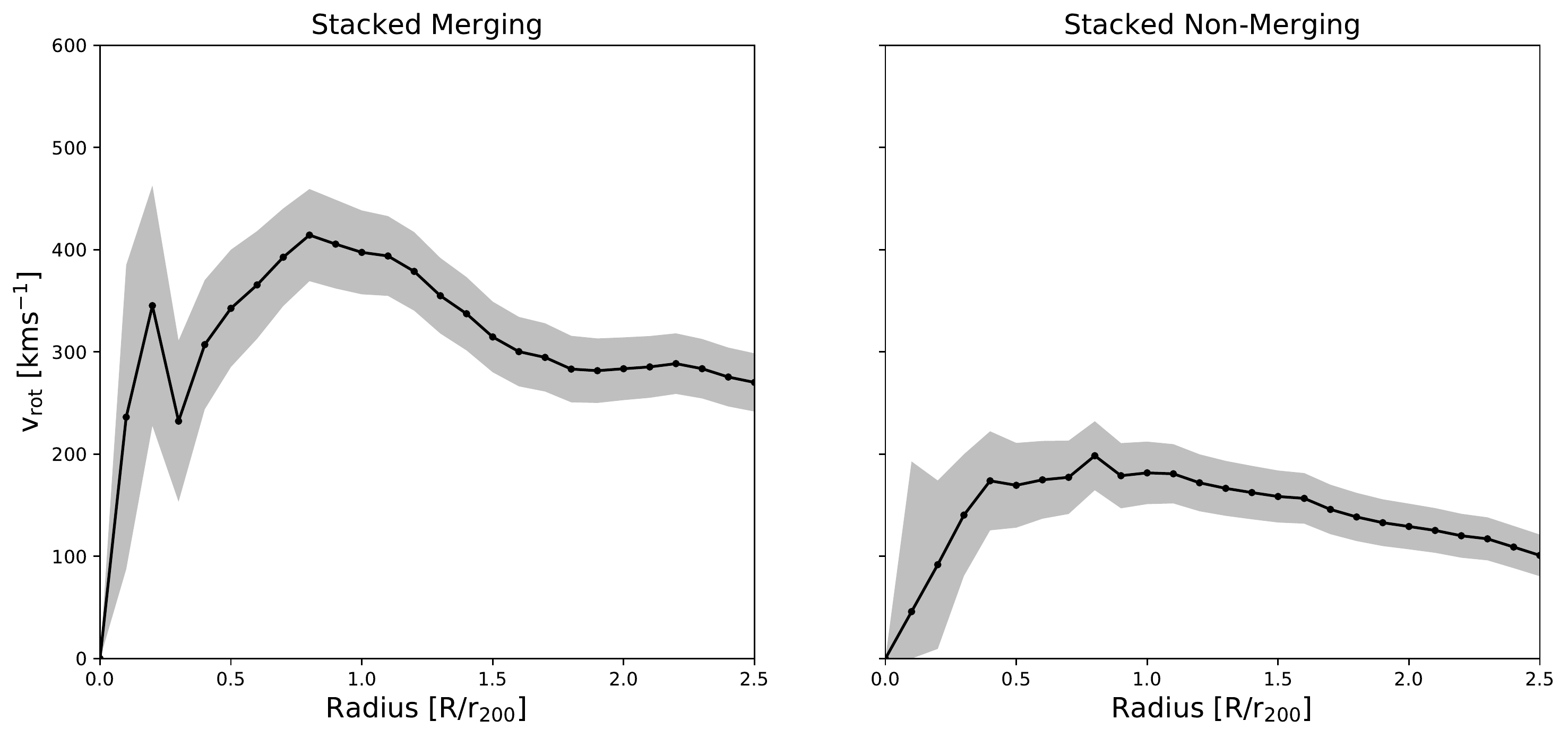}
    \caption{Constrained composite rotational profiles, similarly to Figure \ref{fig:cvrot}, with only BAX clusters lying within redshifts of $0.03\leq z<0.09$ and possessing X-ray luminosities in the range $<6\times10^{44}$erg s$^{-1}$. The overall shape of each of the profiles is retained with some shifts in the magnitude of $v_{\text{rot}}$ across both stacks. The shaded regions represent the uncertainties derived from the propagated standard error as denoted in Equation \ref{eq:unc}.}
    \label{fig:cpvrtcon}
\end{figure*}

\begin{figure*}
    \centering
    \includegraphics[width=0.85\paperwidth]{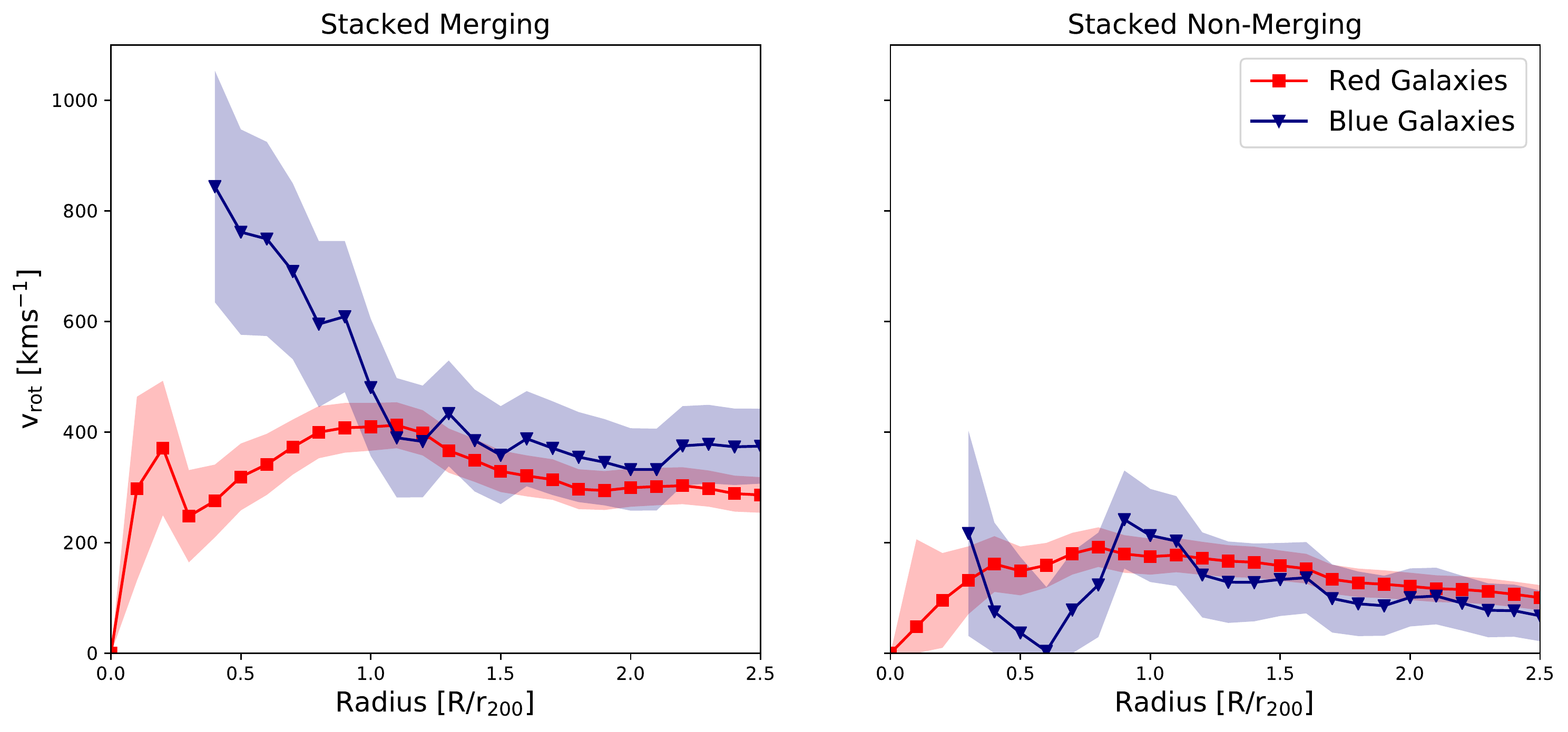}
    \caption{Constrained colour composite rotational profiles, similarly to Figure \ref{fig:cvrotcl}, with only BAX clusters lying within redshifts of $0.03\leq z<0.09$ and possessing X-ray luminosities in the range $<6\times10^{44}$erg s$^{-1}$. Note the differences in $v_{\text{rot}}$ magnitude, especially the blue sub-population in the non-merging composite; loss of signal with a retained shape for values $\lesssim2.0 r_{200}$. The shaded regions represent the uncertainties derived from the propagated standard error as denoted in Equation \ref{eq:unc}.}
    \label{fig:cpvvrtclcon}
\end{figure*}

We show the $L_{X}-z$ distribution for each BAX sub-sample with comparison to the downloaded BAX catalogue in Figure \ref{fig:lxzspl} as a proxy for mass distributions present within our BAX sample. 
Despite some outliers from the non-merging sample, we can see that both the merging and non-merging samples inhabit comparable mass distributions within similar redshifts. 
However, we briefly test how sensitive the rotational profile composites are to the evolutionary epochs and masses by constraining our BAX samples to those that fall within the redshift range of $0.03\leq z<0.09$ and X-ray luminosity values of $<6\times10^{44}$erg s$^{-1}$, this implementation results in a tighter parity between the two sub-samples.
Figures \ref{fig:cpvrtcon} and \ref{fig:cpvvrtclcon} illustrate the constrained sample $v_{\text{rot}}$ composites.

Despite the tighter constraint, the full composites presented in Figure \ref{fig:cpvrtcon} are similar to the unconstrained composite in Figure \ref{fig:cvrot}, with exception of a dampened core in the merging profile and a general shift in the magnitude of $v_{\text{rot}}$ in both merging and non-merging composites.
The most notable difference is found with the constrained colour composite in Figure \ref{fig:cpvvrtclcon} where the blue sub-population of the non-merging stack is subdued with $v_{\text{rot}}$ values falling below the red sub-population within the core regions at $\lesssim r_{200}$.
The constrained non-merging composite suffers large drops in galaxy numbers contributing to the analysis that leaves 721 blue galaxies and 11365 red galaxies.
Although, the general shape of the profile itself is retained with dampened $v_{\text{rot}}$ values as has been consistently shown, with the only notable significant loss found at $\sim r_{200}$ with the peak of the retained \textquoteleft bump' shape of the blue sub-population homogenising with the red sub-population.. 
With this knowledge, alongside the high uncertainties of the non-merging blue sub-population overlapping with the relatively unaffected red sub-population, indicates that the drop in the rotational velocities within the core regions is not significant in displaying a different picture of non-merging systems as shown in Figure \ref{fig:cvrotcl}.
This is aided by considering the large omission of non-merging clusters for the constrained composites, it is therefore, no surprise that the non-merging blue sub-population is more sensitive to the constraints. 
Which is especially the case within the core regions where the number of blue galaxies are fewer. 
Furthermore, it has already been noted how the shape of the blue sub-population in the non-merging sample is consistent, in addition to the overlapping uncertainties, implies that this is predominantly a signal-noise problem.



\section{Simulating the Transverse Motions}
\label{sec:sims}

In this section we describe a 3D simulation of an idealised binary major cluster merger and evaluate how the merger process affects the rotation of the resultant system. We look in particular at how merger phase and viewing angle changes the rotation rate when viewed in an observer-like 2D projection and attempt to draw parallels to the observations.

The simulation is built upon the FLASH Code, a publicly available high performance modular code \citep{Fryxell2000}, utilising the 3D hydrodynamic + N-body capabilities to simulate the gaseous ICM and collisionless dark matter (DM) respectively. Both components being self-gravitating allows the effects of dynamical friction and tidal forces to be captured in the simulation. Taking advantage of the adaptive mesh capabilities of FLASH and refining on particle density results in a maximum resolution in the cluster cores of 19.6 kpc. 

The simulation consists of a 1:2 cluster merger with masses of $5\times10^{14}$ \(\textup{M}_\odot\) and $1\times10^{15}$ \(\textup{M}_\odot\) and $r_{200}$ values of 1672 kpc and 2107 kpc respectively. Following the setup procedure described in \cite{Zuhone2011}, both clusters are non-rotating cool core clusters possessing spherically symmetric single Hernquist mass profiles \citep{Hernquist1990} with a $\beta$-profile for the ICM density. The bulk of the mass is provided by 3 million and 6 million DM particles in the smaller and larger cluster respectively. The initial conditions are set such that at the point the two $r_{200}$ cross one another the relative cluster velocity is 1.1$V_{\mathrm{c}}$ (where $V_{\mathrm{c}} = \sqrt{GM_{\mathrm{vir}}/r_{\mathrm{vir}}} $ ), in accordance with the average infall velocity onto a cluster found from cosmological simulations by \cite{Tormen1997} and \cite{Vitvitska2002}. Following \cite{Poole2006}, we use a tangential velocity component equal to 0.25$V_{\mathrm{c}}$.

To achieve the aims of this section we make the assumption that the simulation's DM particles possess similar motions to that of the galaxies with in the cluster, given that motions of both are collisionless and only feel the effect of dynamical friction. Galaxies would experience higher dynamical friction than a single DM particle given the particles lower mass, however we find that the difference is negligible in this context.
Making this assumption allows us to treat the DM particles as galaxies and use their line of sight velocities to calculate the radial rotation rate of the cluster using the same method described previously for the observational data.  

\begin{figure*}
    \centering
    \includegraphics[width=0.80\paperwidth]{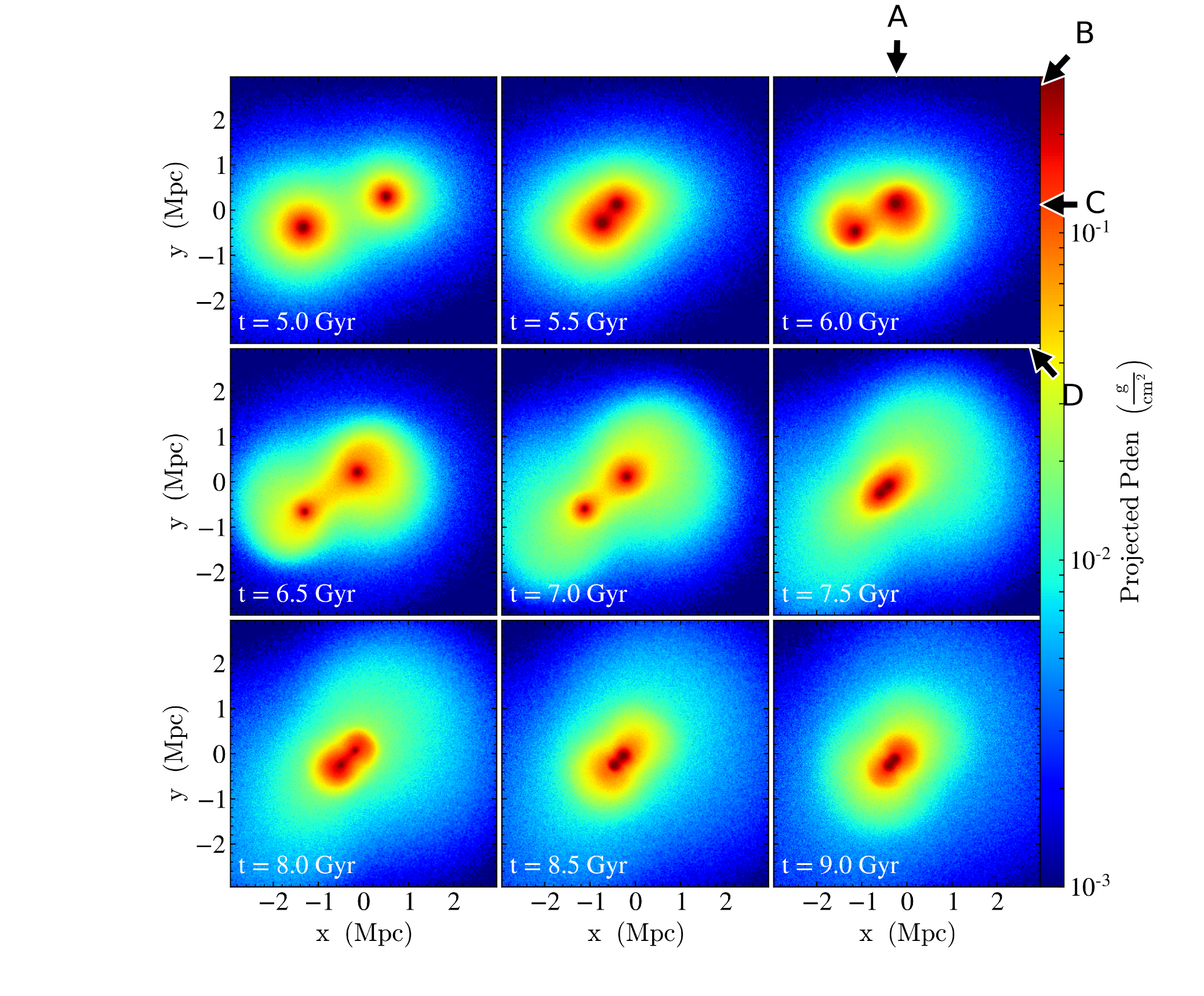}
    \caption{Evolution of the simulated cluster merger shown in projected particle density as a proxy for gravitational potential and line of sight galaxy distribution. Labelled arrows depict the four lines of sights from which the rotation of the system is measured throughout the merger. Line of sight A looks down the y-axis axis. B is down the line of second infall, 37 degrees from A. C is aliened with the x-axis. D is perpendicular the line of second infall. All lines of sight are perpendicular the global axis of rotation and are centred on maximum density.}
    \label{3x3}
\end{figure*}

Figure \ref{3x3} shows the evolution of projected DM in the simulation. From this we can see that after first core passage each cluster core looses all of its tangential velocity relative to the second cluster. This is as a result of the significant dynamical friction the two cores experience traversing one another. Consequently, all future infalls of the two cores proceed along a straight path that links the two first apocentres. This linear motion has the property of always being perpendicular to the axis of rotation of the merger. The cluster cores oscillate for roughly 2.5 Gyrs after second core passage in which time the pass through one another 6 times, after which they become  indistinguishable from one another and hence have merged. If we make the assumption that the BCGs of each cluster remain at the bottom of their respective potential wells throughout the merger then the previous statements regarding the motions of cluster cores can also be considered true for the dumbbell BCGs.


\begin{figure*}
    \centering
    \includegraphics[width=0.80\paperwidth]{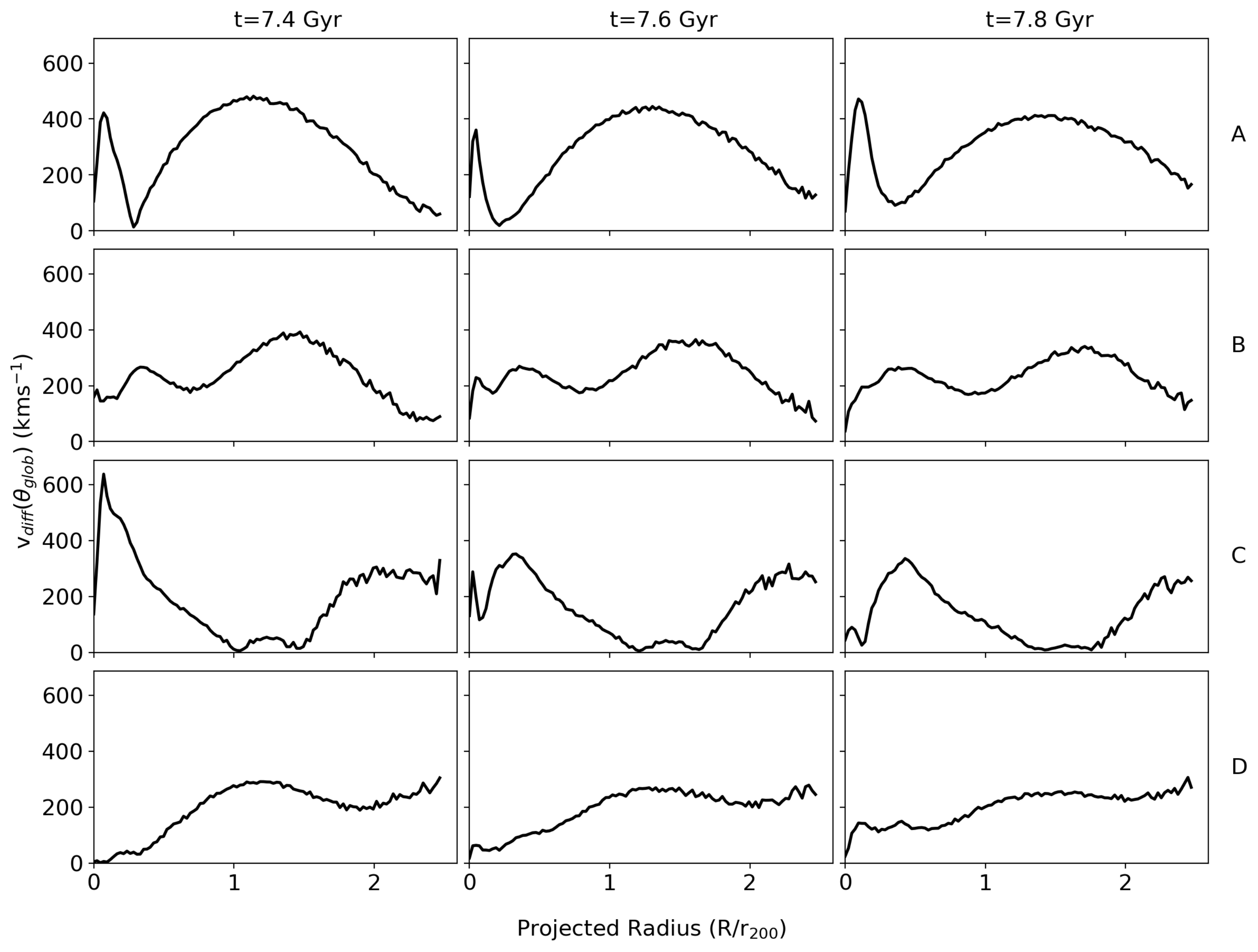}
    \caption{Rotational profiles throughout second core passage (7.4 - 7.8 Gyr) from different viewing angles. Radius is normalised to that of the $r_{200}$ of the more massive cluster. Top row shows the evolution for the line of sight down A (as depicted in Figure \ref{3x3}) and the second, third and fourth rows are down B. C and D respectively. B and D which are parallel and perpendicular to the line of second infall display relatively continuous profiles where as A and C show significant variation particularly within the core.}
    \label{fig:rotation_evolution_multi}
\end{figure*}

Figure \ref{fig:rotation_evolution_multi} shows the radial rotation for the cluster throughout, second passage, from different viewing angles. From this we see that changing the line of sight significantly alters the effect that second infall has on the measured radial rotation rate of the system. Those viewing angles offset from the linear motion of the cluster cores/dumbbell BCGs (A and C in Figure \ref{fig:rotation_evolution_multi}) display dramatic changes in the rotation, dropping from maximum to minimum values within one $r_{200}$ and then to increase again at larger radii. Conversely those parallel or perpendicular (B or D in Figure \ref{fig:rotation_evolution_multi} respectively) posses far more consistent profiles.
The reason for these differences is how the linear motion of the two cores is interpreted via our method of measuring rotation through line of sight motion.

If the merger is viewed such that the line of sight is parallel to the linear motion of the dumbbell BCGs (as with row B) then in the bulk linear motion will average to zero due to the symmetry of the overlapping cores, resulting in it not contributing to the rotation profile.
Similarly, a line of sight perpendicular to the linear dumbbell motion would be unable to detect the bulk velocities of the cores due to no fraction of their motion being down the line of sight, this again results in no \textquoteleft peculiar' increase in rotation whilst still observing the cluster rotation (provided the line of sight was not parallel to the rotation axis where the rotation would not be observable) as can be seen in row D in Figure \ref{fig:rotation_evolution_multi}.

For mergers viewed with an offset from that of the linear motion of the cluster cores/dumbbell BCGs, such as with rows A and C in Figure \ref{fig:rotation_evolution_multi}, the linear motion of the cores is incorporated into the radial rotation. This is due to a substantial component of the velocity of the linear motion being along the line of sight, along with the lack of symmetry the projected system possesses, thus resulting in a fraction of the relative velocities of the cores being interpreted as rotation. This manifests itself as high \textquoteleft peculiar' rotation rates at lower radii. 

\begin{figure}
    \centering
    \includegraphics[width=\columnwidth]{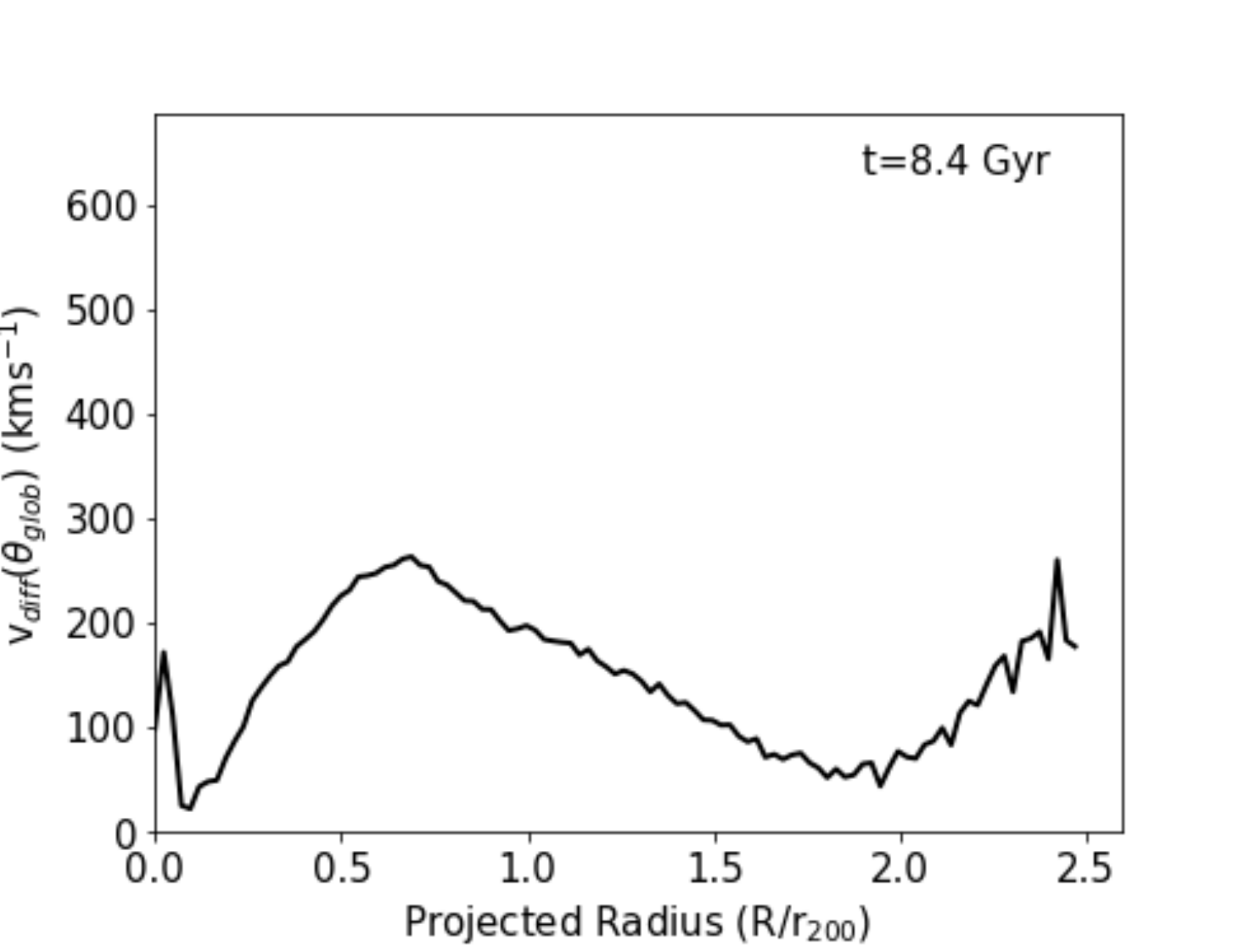}
    \caption{Rotational profiles for cluster merger during third core passage from line of sight C (as depicted in Figure \ref{3x3}). The effect of the linear infall on the rotation of the cluster has become negligible, contributing only a $\sim$170 kms$^{-1}$ increase in within the very core of the system.}
    \label{fig:single}
\end{figure}

Rows A and B of Figure \ref{fig:rotation_evolution_multi} bear resemblance to the rotation of the dumbbell BCG clusters shown in Figures \ref{fig:3391} and \ref{fig:3716}, with their rotation rates being high at low radii, but with rapid reductions to minimum values around 0.5$r_{200}$ and $r_{200}$, then finally increasing again at larger radii. 
The simulation is also in agreement with the prediction for Abell 3391's merger phase drawn from the observations. Based upon observed properties such as angular separation of the cores, rotation profile and the level of sub-structuring, it has been concluded that it is in \textquoteleft post-initial merger phase' as mentioned above. The simulation supports this conclusion as it is only possible to create a rotation profile similar to that of Abell 3391 during second infall of the clusters (shortly before, during and shortly after second core passage). Beyond this the bulk linear motions of the cores, although detectable, is only a minor component in the rotation profile, as can be seen in Figure \ref{fig:single}. Thus comparing Abell 3391 to the simulation suggests that its BCGs are well into their second infall but have not reached second apocentre. This conclusion is in agreement with that made from the observations, however it further constrains what stage the merger has progressed to.

Reverse engineering the previous section gives a framework that could assist observers to further identifying what merger phase a system is in and what angle the observations are being made from.
Systems in which dumbbell BCGs are observed suggest an active mergers phase where the two potential wells can have made up to 6 passages through one another.
A system that displays \textquoteleft peculiar' central velocities along with dumbbell BCGs suggests very early phase mergers, in which the central potentials (BCGs) are on their second infall, i.e. shortly before or shortly after second passage.  It also means the direction of observation is not perpendicular or parallel to the direction of motion of the dumbbell BCDs, neither is it parallel to the axis of rotation. 

This simulation shows that if the viewing angle is favourable then the second infall of a major merger event, during which dumbbell BCGs would be observable, the linear re-infall of central galaxies creates dramatic changes to the observed radial rotation similar to those seen in the rotational profiles of the dumbbell BCG clusters in Figures \ref{fig:3391} and \ref{fig:3716}. However it also shows that we should not expect such \textquoteleft peculiar' rotation in all dumbbell BCG clusters. This is due, in part, to the phase that creates these rotations being relatively short lived ($\sim$500 Myr) when  compared to the time period in which a dumbbell phase could be observed($\sim$2 Gyr). In addition to this, even if the observation was within the 500 Myr window, any viewing angles aliened perpendicular or parallel to the linear motion of the re-infall are unable to detect the \textquoteleft peculiar' rotation.

\section{Discussion \& Summary}
\label{sec:smmy}

Despite the obvious caveat in the disparity between our merging and non-merging cluster sample sizes in this work, they still help to provide a consistency in our current understanding on the formation and evolution of galaxies within different cluster environments found in previous works.
For example, the observed \textquoteleft mixing' of the red and blue sub-populations of galaxies we see in Figure \ref{fig:cvrotcl} corresponds to rising velocity dispersion profiles of mixed red and blue ellipticals found in \cite{Bilton2018}; Mixing of sub-populations kinematically suggests sub-structured pre-processed galaxies are on infall as a result of galaxy-galaxy interactions \citep{Carvalho2017} either prior, or during, off-axial mergers between two sub-clusters; the pronounced population of late-type galaxies on infall in merging environments as inferred by the blue sub-populations of galaxies gaining angular momentum $\lesssim r_{200}$, chiefly thought to be the result of galaxies with spiral morphology that have survived pre-processing (e.g. \citealt{Cava2017,Costa2018,Bilton2018,Nascimento2019}).
The study of rotational profiles would have been aided by the addition of understanding how different morphological sub-populations of cluster galaxies contributed to each of the colour profiles.
However, due to the limitations on resolving such features for every DR8 galaxy, no meaningful analysis could be conducted via the methodology we use within this work.

A common problem with observational studies of galaxy clusters is the limitation of the apparent 2D plane of sky and trying to ascertain information projected onto that sky.
This inherently leads to projection effects due to our inability as observers to comprehend the precise angular and radial separations, therefore, determining the true direction of the rotational axis is not trivial.
The main problem is trying to isolate the true mechanisms behind the observations we record in this work utilising MP17's methods.
All current observational techniques (e.g. \citealt{Kalinkov2005,Hwang2007,Manolopoulou2017}) all determine a velocity gradient between some sort of observer defined axis.
This observer defined axis within itself can be flawed as a result of our chosen centres, with this caveat in mind, we try and maintain consistency through using literary X-ray centres since they are commonly parallel to a cluster's potential.
Therefore, this could potentially indicate that the various techniques we currently have at our disposal are specious, especially when considering we are trying to infer a variety of peculiar motions in a singular $z$-space.
These are the same issues faced with our delineation between merging and non-merging environments via the \cite{Dressler1988} $\Delta$-Test, where, sub-structure is determined through local $z$-space deviations; does the presence of sub-structure genuinely infer rotation via angular momentum donation, or, is this a mere deceptive emulation due to overlapping sub-structures biased by our limited ability to observe galaxy motion?
Ideally, studies on global cluster rotation should combine and model observations between the ICM and the member galaxies; the collisional ICM leaves behind stronger markers of interaction and rotation than the more random (and therefore noisy) collisionless galaxies, which both operate on different time scales \citep{Roettiger2000}.
Furthermore, studies using the kinetic SZ-effect to simulate and analyse the motions of the ICM have shown that the angular momentum and direction between the ICM and dark matter both correlate significantly to imply dark matter dominance \citep{Baldi2017,Baldi2018}.

Overlapping sub-structures in our projected space at least yields results for interactions of galaxies within the cluster.
However, there is still the possibility of interloping sub-structures from other neighbouring clusters.
This is in spite the use of caustic techniques \citep{Diaferio1997,Diaferio1999} to estimate the mass profiles and membership, as well as removing heavily interloping sub-structures using the \cite{Einasto2001} catalogue.
Some examples of interloping can be found in clusters such as Abell 2061 with possible infalling galaxies via a filament from Abell 2067 \citep{Farnsworth2013}; Abell 2065 is believed to be currently undergoing a merger with evidence of two independent sub-clusters with clear structure due to an unequal core merging event \citep{Chatzikos2006}; Abell 3391 is in relatively close proximity to Abell 3395, with X-ray observations indicating the presence of a filament between the two clusters, highlighting the possibility of potential foreign foreground structures \citep{Sugawara2017}.
Although, the use of the cluster caustics performs reasonably in delineating between cluster and non-cluster members, which offsets the reality of a few stragglers invading our cluster membership.
The antithesis to this problem is that by applying our caustics to the more chaotic merging clusters from our sample we result in eliminating genuine cluster members due to the cluster galaxies gaining kinetic energy and increasing their interactions.
\hfill \break

In this work we have acquired MPA-JHU DR8 galaxies cross-matched with a sample of galaxy clusters as defined by the BAX cluster database to build their membership, which are stacked in accordance with their environments, as determined by the $\Delta$-test for sub-structure \citep{Dressler1988}.
This is complemented by NED galaxies of dumbbell clusters \citep{Gregorini1992,Gregorini1994} to allow for comparisons of the dynamics from more extreme and complex systems.
Finally, we compare our perspective rotation methodology from MP17 between our observational DR8 and NED data against FLASH 3D hydrodynamic and N-body simulations of merging clusters \citep{Fryxell2000}.

\hfill \break
\noindent
The key results are summarised as follows:

\begin{enumerate}[(i)]
    \item Cluster rotation $v_{\text{rot}}$ profiles show consistently high rotation until $\sim r_{200}$ with the merging cluster environments (relaxing clusters), whereas non-merging environments commonly depict low $v_{\text{rot}}$ profiles indicative of relaxed clusters undergoing a reduction in the sloshing of galaxies caused by dynamical friction.
    \item Merging cluster environments in our stack exhibit strong core rotation ($\lesssim 0.5 r_{200}$) by the red galaxy sub-population, inferring a sloshing of evolved galaxies as they relax onto a common potential.
    \item The blue galaxy sub-populations in our merging cluster stack have a high $v_{\text{glob}}$ segregation from the red galaxy sub-population in the core regions ($0.4 \lesssim R/r_{200} \lesssim 1.0 r_{200}$) before homogenising with the red sub-population, this may be a consequence of pre-processed sub-groupings that are on infall.
    \item The presence of multi-core dumbbell BCGs in clusters displaying variable $v_{\text{glob}}$ profiles as a result of large peculiar velocities, in-situ of the cluster's rest frame, is indicative of a recent core merger between two originally independent sub-clusters.
    \item Peculiar rotation velocities in dumbbell BCGs are a result of second infall of core galaxies along a linear trajectory that is not aligned with or perpendicular to the line of sight.
    \item The presence of the peculiar rotation velocities are not obligatory in dumbbell BCG clusters due to phases of the dumbbells existence that do not have significant effects on the profile. in addition there are viewing angles that are incapable of measuring the linear motion as the peculiar rotation.  
\end{enumerate}

\section*{Acknowledgements}

\noindent
We would like to extend our thanks to the anonymous referee for their comments and suggestions during peer review of this work.

\noindent
KAP and ER acknowledge the support of STFC through the University of Hull's Consolidated Grant ST/R000840/1.

\noindent
This research made use of Astropy, a community-developed core Python package for Astronomy (\citealt{Collaboration2013}; \citealt{Collaboration2018}).

\noindent
This research has made use of the X-rays Clusters Database (BAX)
which is operated by the Laboratoire d'Astrophysique de Tarbes-Toulouse (LATT),
under contract with the Centre National d'Etudes Spatiales (CNES). 

\noindent
This research has made use of the NASA/IPAC Extragalactic Database (NED), which is operated by the Jet Propulsion Laboratory, California Institute of Technology, under contract with the National Aeronautics and Space Administration.

\noindent
This research has  made use of the \textquotedblleft K-corrections calculator'' service available at \url{http://kcor.sai.msu.ru/}.

\noindent
Funding for SDSS-III has been provided by the Alfred P. Sloan Foundation, the Participating Institutions, the National Science Foundation, and the U.S. Department of Energy Office of Science. The SDSS-III web site is \url{http://www.sdss3.org/}.
SDSS-III is managed by the Astrophysical Research Consortium for the Participating Institutions of the SDSS-III Collaboration including the University of Arizona, the Brazilian Participation Group, Brookhaven National Laboratory, Carnegie Mellon University, University of Florida, the French Participation Group, the German Participation Group, Harvard University, the Instituto de Astrofisica de Canarias, the Michigan State/Notre Dame/JINA Participation Group, Johns Hopkins University, Lawrence Berkeley National Laboratory, Max Planck Institute for Astrophysics, Max Planck Institute for Extraterrestrial Physics, New Mexico State University, New York University, Ohio State University, Pennsylvania State University, University of Portsmouth, Princeton University, the Spanish Participation Group, University of Tokyo, University of Utah, Vanderbilt University, University of Virginia, University of Washington, and Yale University.




\bibliographystyle{mnras}








\bsp	
\label{lastpage}
\end{document}